\documentclass[onecolumn,letter]{aastex6}

\newcommand\pcc{\;{\rm cm}^{-3}}
\newcommand\Msun{\; M_{\odot}}
\newcommand\kms{\; {\rm km}\;{\rm s}^{-1}}
\newcommand\ergs{\; {\rm erg}\;{\rm s}^{-1}}
\newcommand\erg{\; {\rm erg}}
\newcommand\mh{\; m_{\rm H}}
\newcommand\cm{\;{\rm cm}}
\newcommand\yr{\; {\rm yr}}

\newcommand\Myr{\;{\rm Myr}}

\newcommand\pc{\;{\rm pc}}
\newcommand\kpc{\;{\rm kpc}}

\newcommand\Punit{\pcc\,{\rm K}}
\newcommand\Surf{\Msun\;{\rm pc^{-2}}}

\newcommand\Kel{\;{\rm K}}

\newcommand\simgt{\lower.5ex\hbox{$\; \buildrel > \over \sim \;$}}
\newcommand\simlt{\lower.5ex\hbox{$\; \buildrel < \over \sim \;$}}
\newcommand\pderiv[2]{\frac{\partial {#1}}{\partial {#2}}}

\newcommand\rbrackets[1]{\left({#1}\right)}

\newcommand\divergence[2][\rbrackets]{\nabla \cdot #1{#2}}

\newcommand\vel{\mathbf{v}}

\newcommand\rhat{\hat{\mathbf{r}} }

\newcommand\kbol{k_{\rm B}}
\newcommand\rinit{r_{\rm init}}

\newcommand\ESN{E_{\rm SN}}

\newcommand\rsf{r_{\rm sf}}

\newcommand\tsf{t_{\rm sf}}

\newcommand\prad{p_{\rm b}}

\newcommand\Mcl{M_{\rm cl}} 
\newcommand\Pbub{P_{\rm b}}
\newcommand\Ebub{E_{\rm b}}

\newcommand\Mbub{M_{\rm b}}
\newcommand\vbub{v_{\rm b}}
\newcommand\rbub{r_{\rm b}}
\newcommand\rhot{r_{\rm h}}

\newcommand\Thot{T_{\rm h}}
\newcommand\Mhot{M_{\rm h}}
\newcommand\Pamb{P_0}
\newcommand\rhoamb{\rho_{\rm amb}}
\newcommand\namb{n_{\rm amb}}
\newcommand\navg{n_{\rm avg}}
\newcommand\Mhat{\hat{M}_{\rm h}}
\newcommand\phat{\hat{p}_{\rm b}} 
\newcommand\Ehat{\hat{E}_{\rm h}} 
\newcommand\betah{\beta_{\rm h}}
\newcommand\alphah{\alpha_{\rm h}}

\newcommand\tsfm{t_{\rm sf,m}}
\newcommand\rsfm{r_{\rm sf,m}}

\newcommand\dtsn{\Delta t_{\rm SN}}

\newcommand\model[2]{{\tt n{#1}-t{#2}}}
\newcommand\tH{t_{\rm H}}
\newcommand\tHH{t_{\rm 2H}}

\shorttitle{Superbubbles and Galactic Winds}
\shortauthors{Kim, Ostriker, and Raileanu}

\begin{document}

\title{Superbubbles in the Multiphase ISM and the Loading of Galactic Winds}

\author{Chang-Goo Kim, Eve C. Ostriker, and Roberta Raileanu}

\affil{Department of Astrophysical Sciences, Princeton University, Princeton, NJ 08544, USA}
\email{cgkim@astro.princeton.edu, eco@astro.princeton.edu}

\begin{abstract}

We use numerical simulations to analyze the evolution and properties
of superbubbles (SBs), driven by multiple supernovae (SNe), that propagate
into the two-phase (warm/cold), cloudy interstellar medium (ISM).  We
consider a range of mean background densities $\navg=0.1-10\pcc$ and
intervals between SNe $\dtsn=0.01-1\Myr$, and follow each SB
until the radius reaches $\sim(1-2)H$, where $H$
is the characteristic ISM disk thickness.    
Except for embedded dense clouds, 
each SB is hot until a time $\tsfm$ when the shocked warm gas at the
outer front cools and forms an overdense shell.
Subsequently, diffuse gas in the SB interior
remains at $\Thot\sim10^6-10^7\Kel$ with expansion velocity
$v_{\rm h}\sim10^2-10^3\kms$ (both highest for low $\dtsn$). 
At late times, the warm
shell gas velocities are several 10's to $\sim100\kms$.  While shell velocities are
too low to escape from a massive galaxy, they
are high enough to remove substantial mass
from dwarfs. Dense clouds are also
accelerated, reaching a few to 10's of $\kms$.  We measure
the mass in hot gas per SN, $\Mhat$, and the total radial momentum
of the bubble per SN, $\phat$.  After $\tsfm$, 
$\Mhat\sim10-100\Msun$ (highest for low $\navg$), while 
$\phat\sim0.7-3\times10^5\Msun\kms$ (highest for high $\dtsn$).  If
galactic winds in massive galaxies
are loaded by the hot gas in SBs, we conclude that the
mass-loss rates would generally be lower than star formation rates.
Only if the SN cadence is much higher than typical in
galactic disks, as may occur for nuclear starbursts, SBs can break
out while hot and expel up to 10 times the mass locked up in
stars.  The momentum injection values, $\phat$, are 
consistent with requirements to control
star formation rates in galaxies at observed levels.
\end{abstract}

\section{INTRODUCTION}\label{sec:intro}

Many forms of energy originating in stars contribute to heating the
gaseous interstellar- , circumgalactic- , and intergalactic medium
(ISM, CGM, and IGM), but the inputs from supernovae (SNe) play a
unique role because they are so concentrated in space and time.  This
localized deposition of energy leads, through very strong shocks, to
creation of a hot ``third'' phase of the ISM
\citep{1974ApJ...189L.105C,1977ApJ...218..148M} initially in SN
remnants (SNRs) that are highly overpressured relative to their
environment.  Expansion of SN-heated hot gas communicates momentum to
the surrounding ISM and is crucial to maintaining turbulence
in the warm neutral medium (WNM) and cold neutral medium (CNM) phases \citep{2004RvMP...76..125M}, which
would otherwise rapidly collapse to make stars; it is believed that SN
momentum injection is the most important element in the feedback loop
that controls galactic star formation rates
\citep{2011ApJ...731...41O,2011ApJ...743...25K}.  The hot phase
created by SNe is observed to fill a substantial fraction of
the ISM volume within the scale height of the turbulent
CNM/WNM
\citep[e.g.,][]{1998ApJ...497..759F,2007A&A...463.1227K},
sometimes surrounding small clouds of cooler phases
\citep[as in the Local ISM; e.g.][]{2011ARA&A..49..237F},
while on large scales being itself surrounded by shells of cooler gas 
\citep[as in the Orion-Eridanus  Bubble; e.g.][]{1995A&A...300..903B}.
Because of
its high entropy, hot gas tends to rise to create a disk
corona enveloping the cooler ISM phases \citep{1989ApJ...345..372N}.
Depending on its density, coronal gas may cool and condense into
clouds that fall back to the disk, or remain hot and accelerate as a
galactic wind to join the CGM
\citep{1976ApJ...205..762S,1978ApJ...224..768B,1985Natur.317...44C}.

The space-time concentration of SN energy inputs is further enhanced
by stellar clustering. Massive stars are primarily born in clusters,
and while some are ejected to become runaway O stars, the majority
of core-collapse SNe 
explode in close proximity to each other over a period of several tens
of millions of years.  The combined action of many SNe leads to the
development of an expanding superbubble (SB) with a hot interior
and surrounding swept-up shell of cooled post-shock ISM gas 
\citep{1979ARA&A..17..213M,1981Ap&SS..78..273T}.  Large SBs, which
energetically require contributions from multiple SNe, are ubiquitous
in our Galaxy and our neighbors 
\citep[e.g.][]{1979ApJ...229..533H,1984ApJS...55..585H,1988ARA&A..26..145T,
  2007ApJ...656..928P,2015ApJ...808..111O}.
Evolution of SBs depends on the SN rate and properties of the surrounding
ISM.  In cases with sufficiently many SN events (or frequent SNe), SB
evolution is expected to be analogous to the solutions for wind-driven
bubbles powered by continuous energy injection, either in the
simplified case of a uniform ambient medium
\citep[e.g.,][]{1975ApJ...200L.107C,1977ApJ...218..377W,
  1987ApJ...317..190M}, or taking into account stratification in the
background disk
\citep[e.g.,][]{1986PASJ...38..697T,1988ApJ...324..776M,1992ApJ...388...93K}.

Based on results from direct numerical simulations, an increasingly detailed
understanding of the overall three-phase ISM disk is developing
\citep[e.g.,][]{2004A&A...425..899D,2006ApJ...653.1266J,2012ApJ...750..104H,2015MNRAS.454..238W,2014A&A...570A..81H,2015ApJ...814....4L}.
In recent simulations \citep{2015MNRAS.449.1057G,2015MNRAS.454..238W},
the correlation (or lack thereof) of SNe with high-density gas has
been shown to strongly shape the resulting character of the
three-phase ISM, but the effects of SN clustering has not been
investigated in detail.  Instead, the detailed evolution of SBs has
mostly been studied via focused numerical models, in which the
background ISM is treated in a simplified manner.  Continuous thermal
energy injection to a central region has been adopted for most SB
simulations
\citep[e.g.,][]{1999ApJ...513..142M,2000MNRAS.314..511S,2008ApJ...674..157C,2016ApJ...821....7T}, although recently simulations allowing for discrete SN events
have been considered in both spherical symmetry
\citep{2014MNRAS.443.3463S,2016arXiv160601242G}
and for the fully three-dimensional case \citep{2016arXiv160300815Y}. 

The realistic ISM has very large density (and temperature) contrasts,
due to multiphase thermal structure and/or supersonic turbulence. For
single SNR events, the effect of non-uniform background states on SNR
evolution and outcomes has been addressed by several recent direct
numerical simulations.  To model SNR interactions with molecular
clouds, \citet{2015A&A...576A..95I} took as their background state
cold clouds that have been seeded and evolved with supersonic
turbulence, while \citet{2015MNRAS.451.2757W} and
\citet{2015MNRAS.450..504M} adopted background states with an imposed
distribution of density.  In \citet{2015ApJ...802...99K} (hereafter
KO15), we adopted a background state of a cloudy two-phase ISM that
develops from nonlinear saturation of thermal instability.  One of the
main conclusions of these recent studies is that the total radial
momentum injected into the CNM and WNM by the SNR expansion from an
individual SN explosion is
insensitive to the mean background density and largely independent of the 
details of the ambient density distribution.  In KO15, we also considered
a few cases of multiple SNe, and found that the momentum injection per SN is
slightly reduced, but it is still a weak (even weaker) function of the
background density.  This mean momentum per SN, $p_*$, is a key parameter
for turbulence driving and in the theory of self-regulation of star
formation.  The level of $p_*$ obtained in these recent simulations can
explain observations of the turbulent pressure and surface density of
star formation $\Sigma_{\rm SFR}$ in a wide range of galaxies
\citep{2010ApJ...721..975O,2011ApJ...731...41O,2011ApJ...743...25K,2012ApJ...754....2S,2013ApJ...776....1K,2015ApJ...815...67K}.  In this work, we shall
evaluate the momentum injection per SN for situations with multiple SNe,
using a similar numerical setup to that in KO15.

An issue of much interest in both analytic and numerical models of SBs
has been the conditions that enable a SB to break out of the
``ambient'' ISM disk into the galactic halo while still
remaining overpressured relative to the environment \citep[e.g.][]{1988ApJ...324..776M,1989ApJ...337..141M,1992ApJ...388...93K,1999ApJ...516..843B}.  The original
motivation for this question is that overpressured breakout
and Rayleigh-Taylor instability
was considered necessary for releasing hot gas into the galactic
corona, where it could potentially launch a wind.  However, in the
modern understanding of the three-phase, turbulent ISM, there are many
pre-existing low-density channels through which hot gas can vent from
the disk even if a SB is not powerful enough to remain intact until it
reaches the disk scale height.  Thus, even if bubble expansion stalled and there
were no immediate escape routes for hot gas, its high entropy would make
it buoyant.  Although in this paper we do not directly model disk
stratification, we shall discuss various conditions for SB breakout.

An important parameter in analytic and semi-analytic models of
SN-driven galactic winds
\citep[e.g.][]{1985Natur.317...44C,1995ApJ...444..590W,2016MNRAS.455.1830T,2015arXiv150907130B}
is the mass of hot gas launched in the wind per SN.  An alternative
parameterization is in terms of the ``mass loading factor''
$\betah$,
the ratio between the mass of hot gas launched in the wind and the
mass of gas that has (by assumption) collapsed to form stars,
including progenitors of the SNe that drive the wind.  Here, we shall
evaluate the evolution of the mass of hot gas per SN in the interior
of a SB.  The value of this quantity at the time the SB radius is
comparable to the disk scale height allows us to obtain an upper limit
on the mass loading in a galactic wind arising from a region with
certain ISM conditions and SN rate.  Another quantity that is often
used to parameterize SN-driven winds is the energy loading (per
SN or per unit mass of stars formed).  As this is primarily used in
combination with the mass loading to compute the specific enthalpy,
here we will instead measure the temperature of the hot medium within
the SB.  This would represent the typical temperature of the hot ISM phase,
and as it is proportional to the specific enthalpy, it can be used to
constrain the asymptotic wind velocity (assuming adiabatic expansion
such that the Bernoulli parameter is conserved along streamlines).  

In this paper, we extend the previous simulations of KO15 for a more
extensive investigation of SB evolution driven by multiple (discrete)
SN events in the two-phase warm/cold cloudy ISM.  We shall show that,
similar to the situation for individual SNRs, a key stage in the
evolution is when a blastwave propagating into volume-filling warm ISM
first cools, leading to shell formation.  The shell formation time
depends on both the ambient medium density and SN interval (or mass of
star cluster).  We shall show that the SN interval must be smaller
than the shell formation time for the early SB evolution to agree with
the ``continuous energy injection'' limit.

Although we carry out simulations in an unstratified medium, we shall
connect to loading of winds by quantifying the properties of
SBs when their radii are comparable to the scale height of an ISM
disk with the same midplane density as the mean ambient density
in the model.  We shall measure three key quantities in each
simulation at this stage of evolution: the momentum per SN, the mass
of hot gas per SN, and the temperature of the hot gas.  We also evaluate
the distribution of SB mass with velocity at this time.

The plan of this paper is as follows:  In Section \ref{sec:theory} we
review theory of adiabatic SB expansion, and provide reference values for
the expected shell formation time and related quantities.  We also discuss
the analytic theory of SB breakout.  In Section \ref{sec:method} we summarize
the numerical methods and models we use for our simulations.  Section
\ref{sec:result} presents the results of our numerical SB simulations and
analyses, and Section \ref{sec:windload} discusses the implications of
these results for wind loading.  We summarize our conclusions in Section
\ref{sec:sumndis}.  We provide an Appendix to show convergence (as a
function of resolution) in SB properties, and to demonstrate that
SB evolution is independent of the method for injecting SN energy.

\section{ANALYTIC THEORY}\label{sec:theory}

In this section, we reformulate the classical solution for SB
evolution driven by continuous energy injection \citep{1987ApJ...317..190M},
in which the physical
properties of the SB were written in terms of the mechanical luminosity (or
power) or number of SNe.  These solutions are based on the
analogous solutions for wind-blown interstellar bubbles
\citep[e.g.,][]{1972SvA....15..708A,1975ApJ...200L.107C,1977ApJ...218..377W}.
Here, we instead parameterize the power
in terms of mean time interval between SNe, $\dtsn$.

We consider a SB driven by SN explosions originating in a star cluster
with total mass $\Mcl$.  For $\Mcl \simgt 10^3 \Msun$ such that the IMF is
fully sampled, the expected number of SNe is $N_{\rm SN} = \Mcl/m_*$, where
$m_*$ is the total mass of stars formed per SN. For a Kroupa IMF
\citep{2001MNRAS.322..231K}, $m_*\sim100\Msun$.  The SN rate is relatively
constant from $\sim 3\Myr$ to $t_{\rm life}\sim40\Myr$,
\citep[e.g.,][]{1999ApJS..123....3L}, so that
\begin{equation}\label{eq:dtsn}
\dtsn = \frac{t_{\rm life}}{N_{\rm SN}} = 0.4\Myr\; M_{\rm cl,4}^{-1},
\end{equation}
where $M_{\rm cl,4}\equiv \Mcl/10^4\Msun$.  With an energy per SN explosion
$E_{\rm SN}=10^{51}E_{51}\erg $, the total energy that has been injected to
the bubble at time $t$ is 
\begin{equation}\label{eq:sb_energy}
E_{\rm SB}=E_{\rm SN}\frac{t}{\dtsn},
\end{equation}
and the mean power delivered by multiple SNe is given by
\begin{equation}\label{eq:sb_power}
L_{\rm SB}=\dot{E}_{\rm SB}=\frac{E_{\rm SN}}{\dtsn} 
=3.2\times10^{37}\ergs\, E_{51} {\dtsn}_{,6}^{-1}
\end{equation}
where ${\dtsn}_{,6}\equiv \dtsn/\Myr$.

\subsection{Early Adiabatic Expansion}\label{sec:adexp}

Successive multiple SN events contribute to the total energy of the SB, while
the total mass is dominated by the material swept up from its environment.
Before radiative losses become significant, the evolution is analogous to the
Sedov-Taylor solution for a single SN, except with a steady increase in the
energy contained within the expanding blast wave.

From dimensional analysis, the expansion velocities within the SB as well as
the sound speeds in the interior will scale with its outer radius $r$ as
$v\propto r/t$, while the mass contained is $M \propto r^3 \rhoamb$ where
$\rhoamb$ is the density of the surrounding medium (treated as uniform); the
total energy contained therefore varies as $E \propto r^5\rhoamb/t^2$. For
constant input power, energy must increase as $E= L_{\rm SB}t= E_{\rm SN}
t/\dtsn$, which yields $r \propto (L_{\rm SB} t^3/\rhoamb)^{1/5} =(E_{\rm
SN}/\rhoamb \dtsn)^{1/5} t^{3/5}$.  A self-similar solution for the internal
structure of the bubble determines the coefficient ($\sim 0.88$ for
$\gamma=5/3$; see \citealt{1977ApJ...218..377W}).  In terms of the ambient
hydrogen number density $\namb=\rhoamb/(1.4m_H)$, the radius of the outer shock of the
SB can be written 
during the adiabatic expansion stage as
\begin{equation}\label{eq:sb_radius}
r_{\rm ad}=60\pc\,
\rbrackets{\frac{E_{51}}{{\dtsn}_{,6}n_{\rm amb,0}}}^{1/5}t_6^{3/5},
\end{equation}
where $t_6\equiv t/\Myr$ and $n_{\rm amb,0}\equiv \namb/(1\pcc)$.

The expansion velocity of the outer SB shock during the adiabatic stage is
\begin{equation}\label{eq:sb_vel}
v_{\rm ad}\equiv\frac{d r_{\rm ad}}{dt} = 35\kms\,
\rbrackets{\frac{E_{51}}{{\dtsn}_{,6}n_{\rm amb,0}}}^{1/5}t_6^{-2/5},
\end{equation}
the total SB mass during the adiabatic stage is 
\begin{equation}\label{eq:sb_mass}
M_{\rm ad}\equiv\frac{4\pi}{3}r_{\rm ad}^3\rhoamb = 3.2\times10^4\Msun\,
\rbrackets{\frac{E_{51}^3 n_{\rm amb,0}^2 }{\Delta t_{\rm SN,6}^{3}}}^{1/5} t_6^{9/5},
\end{equation}
and the total radial momentum of the SB (treating the mass as concentrated near
the outer shock) is
\begin{equation}\label{eq:sb_mom}
p_{\rm ad}\equiv\frac{4\pi}{3}r_{\rm ad}^3\rhoamb v_{\rm ad} = 
1.1\times10^6\Msun\kms \, \rbrackets{\frac{E_{51}^4 n_{\rm amb,0} }{\Delta
t_{\rm SN,6}^{4}}}^{1/5} t_6^{7/5} .
\end{equation}
For this energy-conserving solution, the momentum per SN in the shell is
\begin{equation}\label{eq:sb_mom_1}
\hat{p}_{\rm ad}\equiv  p_{\rm ad}\frac{\dtsn}{t}=
1.1\times10^6\Msun\kms\,
\rbrackets{E_{51}^4 \Delta t_{\rm SN,6} n_{\rm amb,0} }^{1/5} t_6^{2/5}.
\end{equation}

\subsection{Shell Formation and Post-Radiative Evolution}\label{sec:postrad}

As the SB evolves, the outer regions where the density is highest start to cool
radiatively, forming a thin, dense shell.  The shell formation time for a
single SN explosion in a homogeneous medium is (e.g., Eq. 7 in KO15)
\begin{equation}\label{eq:tsf}
\tsf = 4.4\times10^4\yr\, E_{51}^{0.22}n_{\rm amb,0}^{-0.55}.
\end{equation}
For a SB formed from multiple SN explosions, we can estimate the shell
formation time using Equation (\ref{eq:tsf}) with the energy equal to
$E_{\rm SN} \tsf/\dtsn$ \citep[see
also][]{1988ApJ...324..776M,1992ApJ...388...93K}.  This yields
\begin{equation}\label{eq:tsfm}
\tsfm=1.8\times10^4\yr\, 
E_{51}^{0.28}n_{\rm amb,0}^{-0.71} {\Delta  t}_{\rm SN,6}^{-0.28}.
\end{equation}
Note that in order to be self-consistent with the assumption of continuous
energy injection, it is necessary to have had multiple SN events prior to
shell formation, i.e. $\dtsn < \tsfm$.  Only cases with sufficiently
short SN interval and/or low ambient density,
${\Delta  t}_{\rm SN,6} n_{\rm amb,0}^{0.55}<0.044 E_{51}^{0.22}$, satisfy
this requirement. For cases that do not meet this requirement, shell formation
occurs at the time given in Equation (\ref{eq:tsf}) for a single SN,
when the radius is $r_{\rm sf} = 22.6\pc E_{51}^{0.29} n_{\rm amb,0}^{-0.42}$
(e.g., Eq. 8 in KO15).

Inserting in Equations (\ref{eq:sb_radius}) and (\ref{eq:sb_mom_1}), the
corresponding radius and the momentum injection per SN at the time of shell
formation, for multiple SNe in the continuous energy input limit, are
\begin{equation}\label{eq:rsfm}
  \rsfm \equiv r_{\rm ad}(\tsfm)= 5.5 \pc \,
E_{51}^{0.37}n_{\rm amb,0}^{-0.62} {\Delta  t}_{\rm SN,6}^{-0.37}
\end{equation}
and
\begin{equation}\label{eq:psfm}
\hat{p}_{\rm sf,m} \equiv \hat{p}_{\rm ad}(\tsfm)= 2.3\times10^5\Msun\kms \,
E_{51}^{0.91}n_{\rm amb,0}^{-0.082} {\Delta  t}_{\rm SN,6}^{0.087}.
\end{equation}
This is quite similar to the momentum in the remnant from a single SN at
shell formation,
$p_{\rm sf}=2.2 \times 10^5 \Msun\kms \,E_{51}^{0.93}n_{\rm amb,0}^{-0.13}$
(e.g., Eq. 17 in KO15).

Another interesting quantity is the mass of hot gas in the SB per SN.  Up
to the time of shell formation, the mass of hot gas is just the total mass
of the SB (Equation \ref{eq:sb_mass}); dividing by the number of SNe
at shell formation ($=\tsfm/\dtsn$) yields
\begin{equation}\label{eq:mhsfm}
\hat{M}_{\rm h,sf,m} = 1.3\times10^3\Msun \,
E_{51}^{0.83}n_{\rm amb,0}^{-0.16} {\Delta  t}_{\rm SN,6}^{0.17}.
\end{equation}
Note that, similar to the situation for the mass at shell formation in a single
SNR (e.g., Eq. 11 in KO15), 
this is insensitive to the ambient density, and it is also insensitive to the SN interval.

After shell formation, the low-density interior of the SB remains hot and is
overpressured relative to the ambient medium.  The classical solution for
post-radiative SB evolution
\citep[e.g.,][]{1977ApJ...218..377W,1987ApJ...317..190M,1992ApJ...388...93K}
is similar to the pressure-driven snowplow stage of the SNR for a single SN.
The expansion of the outer SB shell in this stage is assumed to be described by
the momentum equation,
\begin{equation}\label{eq:shell_mom}
\frac{d}{dt}\rbrackets{M_{\rm shell} \frac{d r}{dt}}=4\pi r^2 P_{\rm hot},
\end{equation}
where $M_{\rm shell}\approx \rhoamb4\pi r^3/3$, the exterior pressure is
treated as negligible, and $P_{\rm hot} = E_{\rm hot}(\gamma -1)/(4 \pi r^3/3)$
treating the interior as uniform.
Under the assumption that the interior energy is reduced by 
adiabatic expansion but suffers no radiative losses, the energy equation of the
interior hot gas would be
\begin{equation}\label{eq:shell_energy}
\frac{d E_{\rm hot}}{dt}=L_{\rm SB} 
- 4\pi r^2 P_{\rm hot}\frac{d r}{dt}.
\end{equation}
With $\gamma=5/3$, this again yields $r\propto (L_{\rm SB}
t^3/\rhoamb)^{1/5}$ as in Equation (\ref{eq:sb_radius}).
In contrast to the expansion of a single SNR, where there
are distinguishable changes in the exponents ($r\propto t^{2/5}$ for
energy conserving and $t^{2/7}$ for pressure-driven snowplow), the radius of
the bubble in both the energy conserving and the pressure-driven snowplow
phases have identical parameter dependence, with only
slightly different coefficients (0.88 for the former and 0.76 for the latter).
Thus, the SB radius would follow
\begin{equation}\label{eq:pds_radius}
r_{\rm pds}=52\pc\,
\rbrackets{\frac{E_{51}}{{\dtsn}_{,6}n_{\rm amb,0}}}^{1/5}t_6^{3/5},
\end{equation}
for the pressure-driven snowplow solution \citep{1977ApJ...218..377W}; 
the shell velocity would be a factor of 0.86 below that in Equation
(\ref{eq:sb_vel}), and the shell momentum would be a factor of
0.56 below that in Equation (\ref{eq:sb_mom}).

In practice, the assumptions adopted in the classical pressure-driven SB
evolution are not satisfied in the real ISM.   For the continuous energy
injection model, it is assumed that the hot interior of the bubble is
separated from the cooled shell by a contact discontinuity with continuous
velocity. If, however, the
shell expands at lower velocity than the hot interior, the separation
between the high-velocity, hot interior and the low-velocity, cooled shell is
instead mediated by shocks and/or cooling condensation layers.  The latter
situation occurs after shell formation in the expansion of the remnant
from a single SN \citep[e.g.][KO15]{1988ApJ...334..252C}.  For SBs
driven by small clusters with large $\dtsn$, the evolution may then
resemble a succession of individual SNe more than the continuous limit.

More generally, the high degree of inhomogeneity of the real ISM,
combined with the development of hydrodynamic instabilities
\citep[e.g.,][]{1994ApJ...428..186V,1983ApJ...274..152V}, breaks the
spherical symmetry assumed in the classical solution, such that the interface
between the cooled shell and the hot interior will not be a simple
contact discontinuity. Conduction at interfaces, combined with
turbulent mixing between the dense cooled shell gas and hot interior gas,
enhances radiative losses so that the energy grows more slowly than would be
predicted by Equation (\ref{eq:shell_energy}).  For SB expansion in
the two-phase ISM, energy losses in the hot interior of the SB are
also enhanced by losses from conduction and evaporation of dense
clouds left behind by the expansion of the outer shell in the low-density
intercloud medium; these clouds are also ablated by Kelvin-Helmholtz
unstable interactions with the
surrounding high-velocity hot gas, and mixing into the hot bubble gas
increases its radiative losses.  Recognition of the importance of
these effects has led to intensive numerical investigation, with
dozens of studies focused on the shocked cloud problem alone
\citep[see e.g.][and other citations within]{2015ApJ...805..158S}.

Because we consider expansion of SBs in a cloudy ISM, evolution after
the shell formation stage is far from the classical pressure-driven
bubble solution. Equation (\ref{eq:pds_radius}) therefore
does not describe the realistic post-radiative evolution of the SB radius.  
We thus compare our results only with the early
energy-conserving solutions (Equations \ref{eq:sb_radius},
\ref{eq:sb_mass}, and \ref{eq:sb_mom}), as well as comparing the onset time of
strong cooling to Equation (\ref{eq:tsfm}).

\subsection{Superbubble Breakout}\label{sec:theory_breakout}

Under the assumption that SBs expand as a pressure-driven snowplows (sweeping
up the ambient medium into a cooled shell) with
no radiative cooling in their interior, i.e.
following the generalizations of Equations (\ref{eq:shell_mom}) and
(\ref{eq:shell_energy}) that allow for an external stratified pressure 
and density in the ISM disk (the Kompaneets approximation),
several authors have proposed criteria for SB ``breakout''
from a disk
\citep[e.g.][]{1988ApJ...324..776M,1992ApJ...388...93K,1999ApJ...516..843B}.
Based on Equation (\ref{eq:pds_radius}),
$t_{\rm pds}(H)=H^{5/3}(\rho_{\rm amb} \dtsn/E_{\rm SN})^{1/3}$ is (up to order-unity
factors) the characteristic
timescale for a SB to expand to reach the disk scale height $H$
assuming radiative losses are negligible in the interior.  The 
``breakout'' criterion under this assumption amounts to the requirement that
$t_{\rm pds}(H)$ is sufficiently short (by at least a factor $\sim 3$)
compared to the sound crossing time
over the disk thickness, $\sim H/(P_{\rm amb}/\rhoamb)^{1/2}$.  Physically,
this is also equivalent
to the pressure within the bubble at the time when $r_{\rm pds}=H$
being sufficiently large compared to $P_{\rm amb}$, or the expansion
velocity ($d r_{\rm pds}/dt$) of the shell being sufficiently large compared to
the ambient sound speed.  For an idealized SB with adiabatic interior,
if the breakout criterion 
is satisfied, the shell would accelerate and develop Rayleigh-Taylor instability
as it expands beyond a scale height, whereas otherwise it would stall.

While numerical simulations support the conclusions based on the Kompaneets
approximation analysis for the case of a uniform ambient
medium \citep{1989ApJ...337..141M}, the assumption that the SB interior
remains adiabatic until the radius reaches $\sim H$ is not satisfied for
the realistic cloudy ISM.  As we shall show, while early expansion is
generally consistent with the adiabatic relation of Equation
(\ref{eq:sb_radius}) up to time $\tsfm$,  for $t>\tsfm$ the
SB expands with the total shell momentum (rather than internal energy)
increasing approximately linearly in time.  Also, since realistically
the ambient pressure in
the ISM is generally dominated by the turbulent component rather than the
thermal component, SB shells merge into the turbulent background as their
expansion rates drop, rather than having expansion stalled by
external pressure.

If momentum of the shell grows as
\begin{equation}\label{eq:shell_mom_pdot}
\frac{d}{dt}\rbrackets{M_{\rm shell} \frac{d r}{dt}}=\frac{p_*}{\dtsn}
\end{equation}
for mean momentum per SN $p_*$, then the SB radius will follow a ``momentum
driven snowplow'' relation
\begin{equation}\label{eq:mds_radius}
  r_{\rm mds}=
  \rbrackets{\frac{3p_{*}}{{\dtsn} 2 \pi \rho_{\rm amb}}}^{1/4}t^{1/2}
  = 34\pc\,\rbrackets{\frac{p_{*,5}}{{\dtsn}_{,6}n_{\rm amb,0}}}^{1/4}t_6^{1/2},
\end{equation}
where $p_{*,5}\equiv p_*/(10^5 \kms \Msun)$.
As a function of shell radius, the SB shell velocity in the momentum-driven
limit is
\begin{equation}\label{eq:mds_velocity}
  v_{\rm mds}=
  \rbrackets{\frac{3p_{*}}{{\dtsn} 8 \pi \rho_{\rm amb}}}^{1/2}r^{-1}
  = 5.8\kms\,\rbrackets{\frac{p_{*,5}}{{\dtsn}_{,6}n_{\rm amb,0}}}^{1/2}r_2^{-1},
\end{equation}
where $r_2 \equiv r/10^2\pc$.  
Clear breakout of a SB would correspond to the situation in which
the shell expansion velocity is large enough compared to the
typical velocity dispersion in the disk, $\delta v$, at the time the
shell reaches $\sim H$.  Using Equation (\ref{eq:mds_velocity}) and setting
$r=H$ yields
\begin{equation}\label{eq:vmds_H}
  \frac{v_{\rm mds}(H)}{\delta v } =
  \left(\frac{3 p_*}{\dtsn P_{\rm amb} 8\pi H^2}\right)^{1/2}, 
\end{equation}
where we have substituted $\rho_{\rm amb} \delta v^2 \rightarrow P_{\rm amb}$.
The largest component of $P_{\rm amb}$ is typically the turbulent pressure,
and if the ISM disk overall is consistent with self-regulated
equilibrium with feedback mainly provided by SNe,
$P_{\rm amb} \approx p_* \Sigma_{\rm SFR}/(4 m_*)$ where
$\Sigma_{\rm SFR}$ is the
mean star formation rate per unit area in the disk
\citep{2011ApJ...731...41O,2011ApJ...743...25K}.  Letting
$(\pi H^2 \Sigma_{\rm SFR}/m_*)^{-1} \equiv \Delta t_{\rm SN,H}$ be the mean
interval between SNe within the disk area $\pi H^2$, $v_{\rm mds}/\delta v
=1.2(\Delta t_{\rm SN,H}/\dtsn)^{1/2}$.  If $\dtsn/\Delta t_{\rm SN,H}$
is sufficiently small, the SB shell will remain coherent until breakout
occurs.  For lower-mass clusters with larger $\dtsn$, the shell velocity will
drop below $\delta v$ at an earlier stage, and the SB shell will merge with
the background turbulent ISM structure (which is itself driven by expanding
SNRs and SBs from other SNe).
If multiple clusters within an area $\sim \pi H^2$
act coherently to create a SB, the criterion for visible blowout is
simply that the local star formation rate is sufficiently elevated compared
to its time-averaged value.  

The above considerations imply that the more
massive clusters will create SBs that remain intact until they emerge through
the disk ``surface,'' producing distinctive signatures.
However, even SBs created by lower-mass clusters
with shells that are destroyed 
within the disk may still release hot
overpressured gas that escapes into the galactic halo, 
as we shall discuss in Section \ref{sec:windload}.  There, we shall also
discuss the requirement needed for a SB to break out of the disk prior to
$\tsfm$, i.e. before the onset of strong cooling.

\section{NUMERICAL METHODS \& MODELS}\label{sec:method}

We use the same methods as in KO15.  The inviscid hydrodynamical equations with
optically thin cooling and heating are solved using the {\it Athena} code
\citep{2008ApJS..178..137S,2009NewA...14..139S}.  The mass and momentum
conservation equations are
\begin{equation}\label{eq:cont}
\pderiv{\rho}{t}+\divergence{\rho\vel}=0,
\end{equation}
\begin{equation}\label{eq:mom}
\pderiv{(\rho \vel)}{t}+\divergence{\rho\vel\vel + P} =0,
\end{equation}
and the energy equation, including a source term for net cooling, is
\begin{equation}\label{eq:energy}
\pderiv{E}{t}+\divergence{(E+P)\vel}=-\rho\mathcal{L}.
\end{equation}
The symbols have their usual meanings; $\rho$ is the mass density, $\vel$ is
the velocity, $E\equiv P/(\gamma-1) + \rho v^2/2$ is the total energy density,
$P$ is the gas pressure, and $\gamma=5/3$ is the ratio of specific heats.   The
gas temperature is $T=P/(1.1n_H\kbol)$, where the hydrogen number density is
$n_H=\rho/(1.4\mh)$ for $10\%$ Helium abundance.\footnote{Note that
the temperature for fully ionized gas should be $T=P/(2.3 n_H\kbol)$.
Since we simply fix the mean molecular weight to that of the neutral gas,
however, the temperatures in our simulations are higher than they should be
by factor of $2.3/1.1$ for ionized gas ($T\simgt10^4\Kel$). 
This treatment only causes a slight offset in the  adopted cooling rate,
which depends on
the temperature, but does not affect the sound speed of the gas $c_s^2\equiv
P/\rho$.
}

The net cooling rate per unit volume is $\rho\mathcal{L}\equiv
n_H[n_H\Lambda(T) - \Gamma]$.  We combine cooling functions from
\citet{2002ApJ...564L..97K} and \citet{1993ApJS...88..253S} for low
($T<10^{4.2}\Kel$) and high ($T>10^{4.2}\Kel$) temperature gas, respectively.  A 
constant heating rate per particle $\Gamma$ is only adopted at $T<10^{4.2}\Kel$,
representing photoelectric heating for the CNM and WNM; for hotter gas
$\Gamma =0$.  As we vary the mean density of the ambient medium
from one model to another, we also vary
the heating rate as $\Gamma/\Gamma_0=(n_H/2\pcc)$, where $\Gamma_0 =2\times
10^{-26}\ergs$ is the Solar neighborhood value \citep{2002ApJ...564L..97K}.
This scaling for the heating rate follows the form expected in galactic disks
with self-regulated star formation, in which the photoelectric heating is
approximately proportional to the local star formation rate per unit area, as
well as to the midplane pressure and density 
\citep{2010ApJ...721..975O,2011ApJ...743...25K,2013ApJ...776....1K}.  Explicit
thermal conduction is neglected in this study (see discussion in KO15),
although numerical diffusion at interfaces between hot and cooler phases
can lead to ``evaporation'' from the surface of dense clouds and
energy loss from the hot
medium, similar to the effects of physical conduction. Our convergence
studies are used to assess how these and other resolution-dependent
processes may affect our results.

We study the evolution of SBs produced by multiple SNe in a two-phase medium.
Each SB expands in a ``background'' two-phase medium, which is the result of
nonlinear saturation of the thermal instability in the atomic ISM
\citep{1965ApJ...142..531F}.  This yields CNM clouds
embedded in an intercloud WNM that fills most of volume
($\sim 90\%$).  These phases are in pressure equilibrium, with density and
temperature differing by two orders of magnitude \citep{1995ApJ...443..152W}.   

For each simulation, we represent multiple SNe via successive explosions at the
center of the domain, with fixed time intervals between events.  We consider 9
models with three different values for the mean density of the ambient medium
$n_{\rm avg} =$ 0.1, 1, and $10\pcc$, and three different time intervals
$\dtsn=$ 0.01, 0.1 and 1 Myr.  Each model is named based on these two key
parameters; for example, \model{1}{0.1} denotes $n_{\rm avg}=1\pcc$ and
$\dtsn=0.1\Myr$.  Table~\ref{tbl:models} lists the parameters for each model,
including $\dtsn$ and $n_{\rm avg}$ in Columns (2) and (3), the mean
density of the WNM, $n_w$, and the mean pressure of the background ambient
two-phase medium, $P_0$,  in Columns (4) and (5).  Column (6) lists the spatial
resolution of the simulation, which varies with $n_{\rm avg}$ to satisfy the
consistent convergence condition we determined in KO15: 
$\Delta_x < \rsf/10$, where
$\rsf =22.6\pc\, (n_{\rm avg}/1\pcc)^{-0.42}$ is the predicted shell formation
radius for a single SN explosion.  We have run two additional models for Model
\model{1}{0.1} to confirm numerical convergence (see Appendix
\ref{sec:convergence}).

In Column (7), we list the typical scale height for an ISM disk that has mean
midplane density $n_{\rm avg}$, defined by
\begin{equation}\label{eq:H_def}
H \equiv 104\pc\, (n_{\rm avg}/1\pcc)^{-1/2}.
\end{equation}
This is a rough estimate using vertical dynamical equilibrium,
$ H=\sigma_z [\pi G (\pi \rho_{\rm mid} + 4\rho_{\rm sd})]^{-1/2} $
where the total gas surface density $\Sigma=H\sqrt{2\pi} \rho_{\rm mid}$.  If 
the midplane volume density of stars and dark matter, $\rho_{\rm sd}$, scales
with the midplane gas density $\rho_{\rm mid}$ (or else if the gas density
dominates), this yields $H\propto \rho_{\rm mid}^{-1/2}$.  For the
normalization, we use the results from \citet{2013ApJ...776....1K}, in which we
obtained $H\sim 85\pc$ for the midplane density $n_{\rm avg}\sim 1.5\pcc$ from
self-consistent modeling of galactic disks with feedback from star formation.
In the present simulations, we do not in fact include any vertical gravity, so
that our models are unstratified.  However, it is useful to keep in mind an
approximate value for the scale height, in order to define SB properties at the
time the bubble radius reaches what the warm/cold ISM scale height would be,
and starts to break out into circumgalactic space.

To represent successive SN explosions at an interval $\dtsn$, we assign thermal
energy of $E_{\rm SN}=10^{51}\erg$ within a ``feedback region.''
\footnote{We have confirmed that if, rather than injecting thermal energy,
  we introduce the same amount of kinetic energy as expanding ejecta,
  our results are essentially the same.  See Appendix \ref{sec:convergence}.}
The size of
the feedback region at any time is determined by the largest possible size that
satisfies the convergence condition of KO15, $\rinit<\rsf/3$.  In practice, we
begin by setting $\rinit=3\Delta_x$ and calculate the mean density for the
total gas mass of cells on the grid within $\rinit$ {\em plus} the mass of the ejecta and
circumstellar material $M_{\rm min}=10\Msun$.  We then calculate $\rsf$ for
that density. If $\rinit$ is smaller than $\rsf/3$, we increase $\rinit$ by
$\Delta_x/2$ and iterate until $\rinit$ reaches $\rsf/3$.  The gas mass
density, momentum density, and pressure for each zone within $\rinit$ are
initially reset to the mean values in the feedback region, and we then add
$E_{\rm SN}/V_{\rm init}$ to the internal energy density in each zone,
where the volume of the feedback region is $V_{\rm init}\equiv \sum_{r<\rinit}
\Delta_x^3$.  By including mass for ejecta and circumstellar material, the
density in the feedback region density does not become too small (which would
lead to numerical difficulties).  We have confirmed that the specific value for
$M_{\rm min}$ does not affect any outcomes discussed in this paper, since this
mass is small compared to the swept-up mass, which governs the dynamics of the
SB.

\begin{deluxetable}{lcccccc}
\tablecaption{Model Parameters \label{tbl:models}}
\tablehead{ 
\colhead{Model} &
\colhead{$\Delta t_{SN}$} &
\colhead{$n_{\rm avg}$} &
\colhead{$n_w$} &
\colhead{$\Pamb$} &
\colhead{$\Delta_x$} &
\colhead{$H$} 
}
\colnumbers
\startdata 
\model{0.1}{0.01}  & 0.01 &     &      &                &      &      \\
\model{0.1}{0.1}   & 0.1  & 0.1 & 0.017& 99             & 6    & 329  \\
\model{0.1}{1}     & 1    &     &      &                &      &      \\
\hline                                                               
\model{1}{0.01}    & 0.01 &     &      &                &      &      \\
\model{1}{0.1}     & 0.1  & 1   & 0.14 & $1.1\times10^3$& 3    & 104  \\
\model{1}{1}       & 1    &     &      &                &      &      \\
\hline                                                              
\model{10}{0.01}   & 0.01 &     &      &                &      &      \\
\model{10}{0.1}    & 0.1  & 10  & 1.5  & $1.2\times10^4$& 0.75 & 33   \\
\model{10}{1}      & 1    &     &      &                &      &      \\ 
\hline                                                              
\model{1}{0.1-low} & 0.1  & 1   & 0.14 & $1.1\times10^3$& 6    & 104  \\
\model{1}{0.1-high}& 0.1  & 1   & 0.14 & $1.1\times10^3$& 1.5  & 104  \\
\enddata
\tablecomments{
Col.~(1): model name.
Col.~(2): time interval between SNe, in Myr.
Col.~(3): mean density of the ambient medium, in $\pcc$.
Col.~(4): mean density of the WNM, in $\pcc$.
Col.~(5): mean pressure of the ambient medium, in $k_B\Punit$.
Col.~(6): resolution, in pc.
Col.~(7): reference ISM scale height (see Eq. \ref{eq:H_def}), in pc.
}
\end{deluxetable}

\section{NUMERICAL SIMULATION RESULTS}\label{sec:result}

Before describing the model results, we establish definitions for
separate components of the SB.  First, we define the
``bubble'' component as all the gas that has been affected by the
blast wave. This is comprised of all zones with $T>10^5\Kel$ or
$v>1\kms$.  The ambient medium is comprised of the remainder of zones
in the domain (note that ambient gas is initially stationary, but
small velocities develop since pressure balance between the warm and
cold phases is not perfect). We define the ``hot'' gas as all zones
with $T>10^5\Kel$.  All the hot gas is part of the bubble, but the
bubble also contains gas that has been shocked and then radiatively
cooled below $10^5\Kel$.

We measure the equivalent spherical radius, mass, total energy, pressure, and
temperature of the hot and bubble gas. The radius is $r_{\rm c}\equiv (3V_{\rm
c}/4\pi)^{1/3}$ and $V_{\rm c} \equiv \sum_{\rm c} \Delta_x^3$, where the gas
component `c' can be either `hot' or `bubble', and $\sum_{\rm c}$ is summation
over the zones that satisfy the definition of each gas component.  The mass and
energy are defined by $M_{\rm c} \equiv \sum_{\rm c} \rho\Delta_x^3$ and
$E_{\rm c} \equiv \sum_{\rm c} [P/(\gamma-1)+\rho v^2/2]\Delta_x^3$,
respectively.  The pressure and temperature are defined with volume and
mass-weighted means, respectively, as $P_{\rm c} \equiv \sum_{\rm c}
P\Delta_x^3/V_{\rm c}$ and $T_{\rm c} \equiv 1.27m_HP_{\rm c}V_{\rm c}/(M_{\rm
c}k_B)$.  Finally, the total radial momentum of the bubble is calculated by
$\prad\equiv \sum_{\rm b} \rho \vel\cdot\rhat \Delta_x^3$.

In Table~\ref{tbl:result}, we summarize properties of SB evolution for each
model.  The expected shell formation time $\tsfm$ from Equation (\ref{eq:tsfm})
is listed in Column (2), and the measured times when $\rbub=H$ and $2H$,
$\tH$ and $\tHH$, are listed in Columns (3) and (4), respectively.  We also
list the reference scale height in Column (5) and the measured bubble radius at
$\tsfm$ in Column (6).  The measured bubble mass,
mean velocity $\vbub\equiv \prad/\Mbub$,
and the hot gas temperature in the simulation at
$\tH$ and $\tHH$ are listed in Columns (7)-(12).  As noted above, because we
do not allow the mean molecular weight to vary in the simulation, the true
temperature of the hot medium would be a factor of two lower.

To connect our results to loading of galactic winds, we measure the hot gas
mass and
thermal energy per SN event defined by $\Mhat\equiv \Mhot/N_{\rm SN}$ and
$\Ehat \equiv E_{\rm th,h}/N_{\rm SN}$, respectively, with $N_{\rm SN} =
\lfloor t/\dtsn\rfloor +1$, where $\lfloor x \rfloor$ is the floor function
that maps a real number $x$ to the largest previous integer.  To
connect our results to driving of turbulence within galactic disks, we
measure the total radial momentum of the bubble per SN event as
$\phat \equiv p_{\rm b}/N_{\rm SN}$.
In Table~\ref{tbl:perSN}, we summarize the SB properties per SN
measured at $\tH$ and $\tHH$.

From Table~\ref{tbl:result}, the SB would expand to $H$ within $10^5$-$10^6
\yr$ for the parameter range considered,
but because the expansion slows over time, reaching $2H$ requires
$10^6$-$10^7 \yr$.  Table~\ref{tbl:result} also shows that $\rbub(\tsfm)<H$ by a
large margin for all cases except models with $\dtsn =0.01 \Myr$. As we shall
discuss in Section~\ref{sec:windload}, this implies that unless $\dtsn$ is
quite short, SBs cool before breaking out of the disk.  In turn, this
suggests that substantial hot gas mass loss in SN-driven winds can only occur in
localized regions within galaxies where there are fairly massive clusters, or
where several clusters are in close enough proximity (e.g. in galactic center
regions) such that $\dtsn$ from the combined system is short.  Indeed,
Table~\ref{tbl:perSN} shows that $\Mhat(H) > 100 \Msun$ in only two
cases with $\dtsn =0.01 \Myr$.  This implies that in most cases (for
the present range of parameters), the hot gas
mass in a SB at breakout is less than the total mass in newly formed stars of
the cluster that drove the SB.
However, Table~\ref{tbl:result} also shows
that in essentially all cases, the hot gas has temperature $> 10^6\Kel$ at the
time the SB would break out of the disk.  Thus, while the amount of hot gas
expelled per star formed may not always be large, the sound speed is generally
high enough to drive a wind that can escape the galactic potential well (see
Section~\ref{sec:perSN}).

For most cases, $\Ehat(H)/10^{51}\erg$ is only
a few percent or less, implying that most of the input energy is lost to a
combination of radiative cooling and kinetic energy  in the
warm/cold ISM before SB breakout.
Indeed, $\hat{E}_{\rm b}(H)\gg \Ehat(H)$ in all cases except those where
$\tsfm \sim \tH$.

The mean velocity of the SB substantially exceeds $10\kms$ at $\tH$ for
the models with $\dtsn=0.01$ and $0.1\Myr$.  Since this exceeds typical
background turbulence levels in observed galaxies, it suggests that
SBs would remain coherent in their appearance until breakout for SBs
driven with a high SN cadence, as argued in Section \ref{sec:theory_breakout}.
Cases with $\dtsn=1\Myr$ have lower $\vbub(H)$, suggesting that for lower
mass clusters, the SB shell would instead merge with the background ISM
turbulence before breaking out of the disk.  In all cases, the mean value
of $\vbub(H)$ is smaller than the escape speed of all but very low mass
halos, indicating that the shell would not escape as a whole from most
galaxies.  However, we shall see in Section \ref{sec:pdf} that there is
substantial gas mass with velocities above $50\kms$ for the cases
$\dtsn=0.01$ and $0.1\Myr$, which would be able to escape from dwarf galaxies.

\vbox{
\begin{deluxetable}{lccccccccccc}
\tablecaption{Superbubble Evolution Properties \label{tbl:result} }
\tablehead{ 
\colhead{Model} &
\colhead{$\tsfm$} &
\colhead{$\tH$} & 
\colhead{$\tHH$} &
\colhead{$H$} &
\colhead{$\rbub(\tsfm)$} &
\colhead{$\Mbub (H)$} &
\colhead{$\Mbub (2H)$} & 
\colhead{$\vbub (H)$} &
\colhead{$\vbub (2H)$} & 
\colhead{$\Thot (H)$} &
\colhead{$\Thot (2H)$} 
}
\colnumbers
\startdata 
 \model{0.1}{0.01} & 1.23  & 1.23  &\nodata& 329    & 329     & 3.18   &\nodata &  39.8   & \nodata  &  3.25   & \nodata   \\
 \model{0.1}{0.1}  & 0.65  & 3.59  &\nodata& 329    & 153     & 3.31   &\nodata &  16.9   & \nodata  &  3.10   & \nodata   \\
 \model{0.1}{1}    & 0.34  & 8.32  &\nodata& 329    & 107     & 2.98   &\nodata &   7.87  & \nodata  &  1.50   & \nodata   \\
 \model{1}{0.01}   & 0.28  & 0.32  & 1.69  & 104    &  98     & 0.81   & 8.13   &  46.5   & 3.32     &  7.75   & 10.1      \\
 \model{1}{0.1}    & 0.15  & 1.01  & 4.12  & 104    &  53     & 0.79   & 6.93   &  19.6   & 2.07     &  4.12   &  4.06     \\
 \model{1}{1}      & 0.076 & 1.86  & 8.72  & 104    &  44     & 0.64   & 4.75   &  8.00   & 0.83     &  1.59   &  2.13     \\
 \model{10}{0.01}  & 0.051 & 0.14  & 0.58  &  33    &  24     & 0.26   & 2.26   &  45.2   & 4.60     & 14.0    & 11.4      \\
 \model{10}{0.1}   & 0.027 & 0.25  & 1.24  &  33    &  18     & 0.21   & 1.81   &  21.7   & 1.93     &  5.43   &  4.41     \\
 \model{10}{1}     & 0.014 & 0.28  & 2.05  &  33    &  15     & 0.18   & 1.18   &  12.4   & 1.56     &  0.70   &  1.98     \\
\hline                                                        
\model{1}{0.1-high}& 0.15  & 0.95  & 3.90  & 104    &  52     & 0.76   & 7.04   &  18.8   & 1.93     &  3.99   &  4.00     \\ 
\model{1}{0.1-low} & 0.15  & 1.03  & 4.33  & 104    &  54     & 0.79   & 7.23   &  20.6   & 2.08     &  3.74   &  4.04     \\
\model{1}{0.01-ej} & 0.28  & 0.29  & 1.59  & 104    &  101    & 0.81   & 8.24   &  46.5   & 3.15     &  4.99   &  6.57     \\
\model{1}{0.1-ej}  & 0.15  & 1.00  & 4.03  & 104    &  53     & 0.79   & 7.29   &  16.9   & 1.64     &  6.41   &  4.22     \\
\model{1}{1-ej}    & 0.076 & 1.84  & 8.86  & 104    &  44     & 0.64   & 4.76   &  8.09   & 0.76     &  1.57   &  2.39      
\enddata
\tablecomments{
Col.~(1): model name.
Col.~(2): theoretical shell formation time $\tsfm$ from Eq.(\ref{eq:tsfm}), in Myr.
Cols.~(3-4): times when the SB reaches $\rbub=H$ and $2H$, in Myr. 
Col.~(5): reference disk scale height $H\equiv 104\pc (n_{\rm avg}/1\pcc)^{-1/2}$, in pc.
Col.~(6): measured bubble radius at $\tsfm$, in pc.
Cols.~(7-8): masses of the bubble at $H$ and $2H$, in $10^5\Msun$.
Cols.~(9-10): velocity of the bubble gas at $H$ and $2H$, in $\kms$.
Cols.~(11-12): temperatures of the hot gas at $H$ and $2H$, in $10^6\Kel$.
}
\end{deluxetable}

}

\vbox{

\begin{deluxetable}{lcccccccc}
\tablecaption{Superbubble Properties per SN \label{tbl:perSN}}
\tablehead{ 
\colhead{Model} &
\colhead{$\Mhat (H)$} &
\colhead{$\Mhat (2H)$} & 
\colhead{$\phat (H)$} &
\colhead{$\phat (2H)$} &
\colhead{$\Ehat (H)$} &
\colhead{$\Ehat (2H)$}&
\colhead{$\hat{E}_{\rm b} (H)$}&
\colhead{$\hat{E}_{\rm b} (2H)$}
}
\colnumbers
\startdata 
 \model{0.1}{0.01} & 386    & \nodata & 1.02  &\nodata & 0.15  &\nodata  &  0.38  & \nodata \\
 \model{0.1}{0.1}  &  92    & \nodata & 1.55  &\nodata & 0.034 &\nodata  &  0.16  & \nodata \\
 \model{0.1}{1}    &  48    & \nodata & 2.61  &\nodata & 0.0084&\nodata  &  0.084 & \nodata \\
 \model{1}{0.01}   & 169    &  32     & 1.18  &  0.84  & 0.15  &  0.039  &  0.44  &   0.13  \\
 \model{1}{0.1}    &  51    &  39     & 1.40  &  1.30  & 0.025 &  0.019  &  0.14  &   0.078 \\
 \model{1}{1}      &   4.8  &  11     & 2.57  &  1.98  & 0.0008&  0.0028 &  0.064 &   0.045 \\
 \model{10}{0.01}  &  21    &  16     & 0.84  &  0.74  & 0.034 &  0.022  &  0.21  &   0.10  \\
 \model{10}{0.1}   &   3.9  &   9.0   & 1.49  &  1.16  & 0.0025&  0.0045 &  0.087 &   0.043 \\
 \model{10}{1}     &   1.6  &   0.71  & 2.27  &  1.84  & 0.0001&  0.0002 &  0.083 &   0.054 \\
\hline                                                                   
\model{1}{0.1-high}& 57     & 36      & 1.43   & 1.36  & 0.027 &  0.017  & 0.14   &  0.083 \\ 
\model{1}{0.1-low} & 58     & 38      & 1.49   & 1.37  & 0.025 &  0.018  & 0.15   &  0.078 \\
\model{1}{0.01-ej} & 344    & 38      & 1.26   & 0.87  & 0.20  &  0.030  & 0.51   &  0.12  \\
\model{1}{0.1-ej}  & 34     & 32      & 1.33   & 1.19  & 0.025 &  0.016  & 0.11   &  0.061 \\
\model{1}{1-ej}    & 4.9    & 11      & 2.58   & 1.81  & 0.0008&  0.031  & 0.064  &  0.041
\enddata
\tablecomments{
Col.~(1): model name.
Cols.~(2-3): mass of the hot gas per SN at $H$ and $2H$, in $\Msun$.
Cols.~(4-5): total radial momentum of the bubble per SN at $H$ and $2H$, in units $10^5\Msun \kms$.
Cols.~(6-7): thermal energy of the hot gas per SN at $H$ and $2H$, in units $10^{51} \erg$.
Cols.~(8-9): total bubble energy per SN at $H$ and $2H$, in units $10^{51} \erg$.
}
\end{deluxetable}

}

\subsection{Time Evolution of Overall Bubble Properties}\label{sec:tevol}

\begin{figure}
\plotone{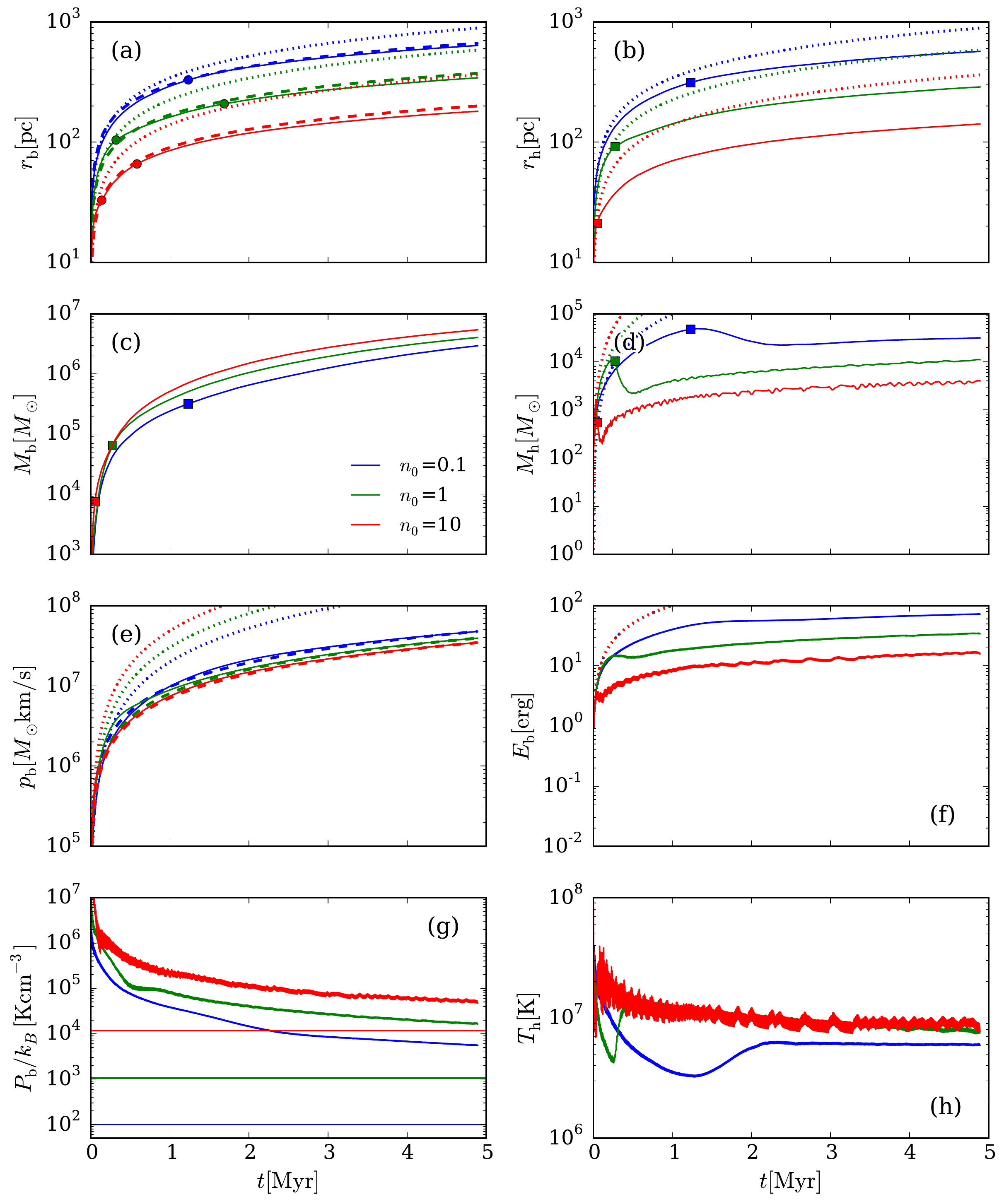}
\caption{Time evolution of the models with $\dtsn= 0.01\Myr$.  Panels show (a)
radius of the bubble $\rbub$, (b) radius of hot gas $\rhot$, (c) mass of the
bubble $\Mbub$, (d) mass of hot gas $\Mhot$, (e) total radial momentum $\prad$,
(f) total energy of the bubble $\Ebub$, (g) pressure of the bubble
$\Pbub/\kbol$, and (h) temperature of hot gas $\Thot$. The circles in panel (a)
indicate the times when the corresponding radii reached $H$ and $2H$, $\tH$ and
$\tHH$, respectively.  The squares in panels (b), (c), and (d) denote the
corresponding values at $t=\tsfm$.  The dotted lines in (a) and (b), (e), and
(f) denote analytic predictions for radius, momentum, and total injected energy
in the energy-conserving continuous limit from
Equations (\ref{eq:sb_radius}), (\ref{eq:sb_mom}), and
(\ref{eq:sb_energy}), respectively,
while the dotted lines in (d) indicate
the warm swept-up masses using the radius predicted from Equation
(\ref{eq:sb_radius}). 
The dashed lines in (a) and (e) denote analytic predictions for radius and momentum
in the momentum driven snowplow stage from Equations (\ref{eq:mds_radius})
and $p_*t/\dtsn$, respectively.
The solid lines in (g) show the ambient medium pressure
for reference.  \label{fig:tevol_t001}}
\end{figure}

\begin{figure}
\plotone{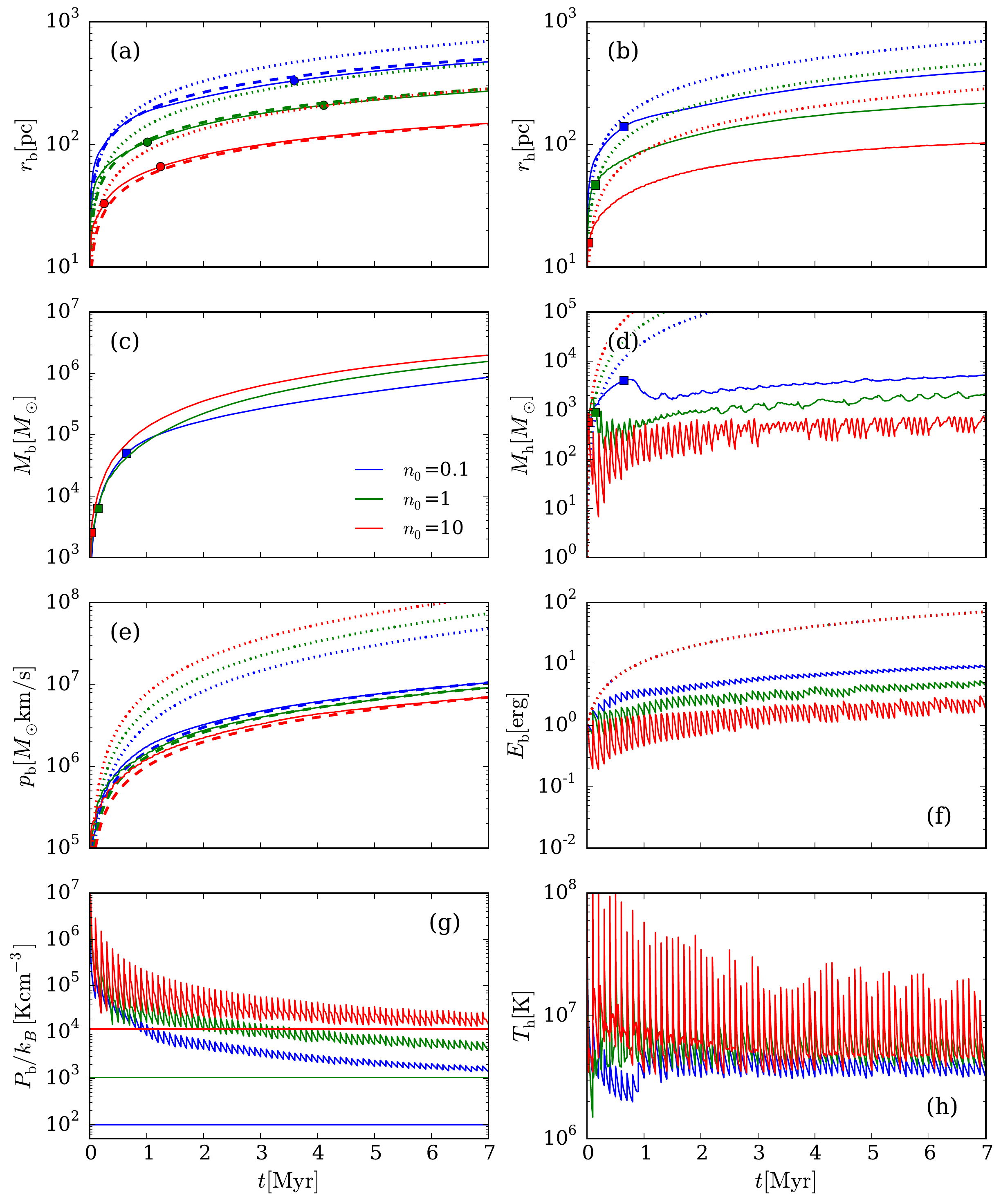}
\caption{Same as Figure~\ref{fig:tevol_t001}, but for models with $\dtsn =
0.1\Myr$.  \label{fig:tevol_t01}}
\end{figure}

\begin{figure}
\plotone{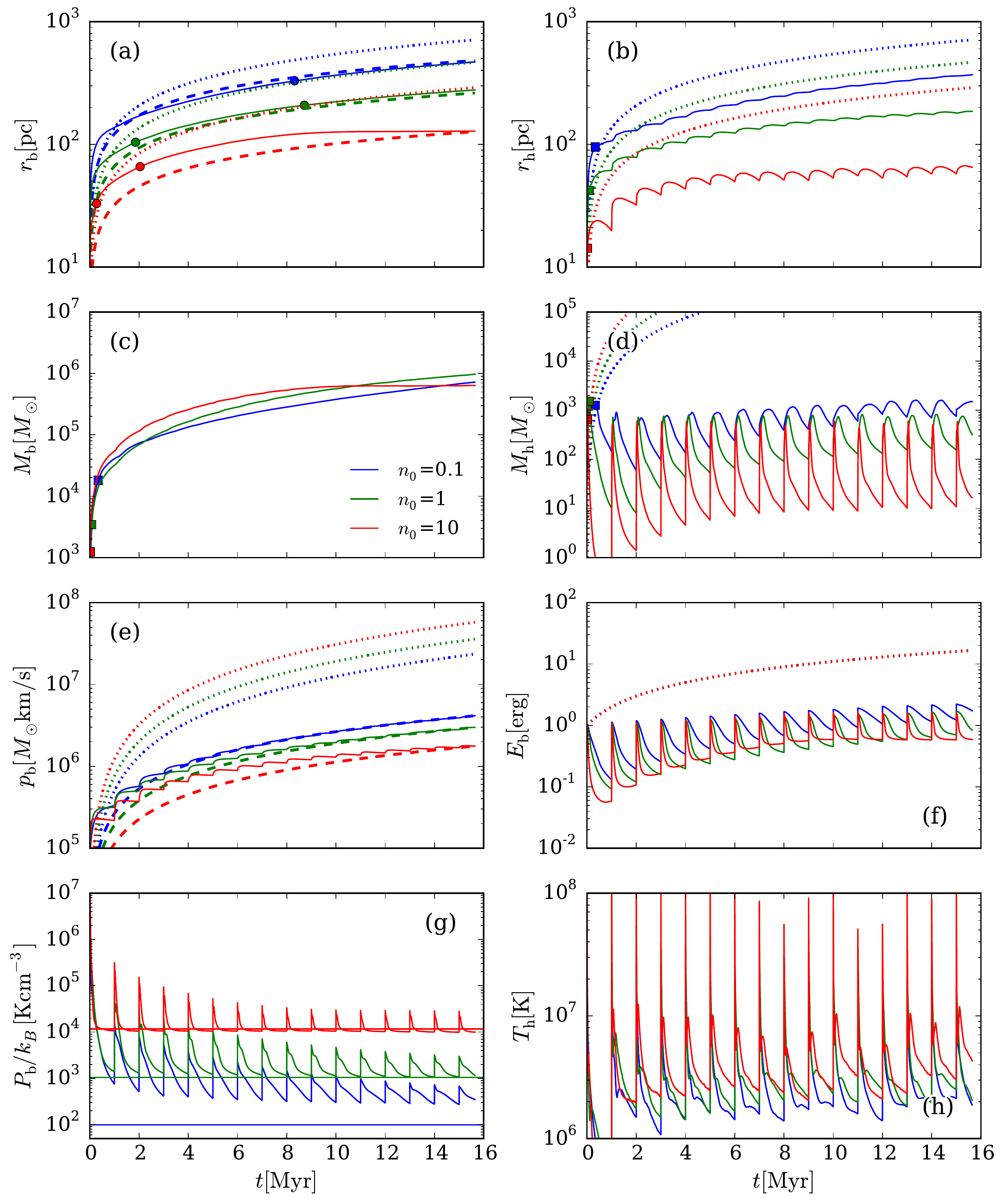}
\caption{Same as Figure~\ref{fig:tevol_t001}, but for models with $\dtsn =
1\Myr$.  \label{fig:tevol_t1}}
\end{figure}

Figures \ref{fig:tevol_t001}-\ref{fig:tevol_t1} plot time evolution of
SB properties for $\dtsn=0.01$, 0.1, and 1 Myr, respectively. 
Each panel shows 
(a) bubble radius $\rbub$; (b) hot gas radius $\rhot$;
(c) bubble mass $\Mbub$; (d) hot gas mass $\Mhot$; (e) bubble momentum
$\prad$; (f) bubble energy $\Ebub$; (g) bubble pressure $\Pbub$; (h)
hot gas temperature $\Thot$.  Analytic predictions of SB radius
(Eq.(\ref{eq:sb_radius})), momentum (Eq.  (\ref{eq:sb_mom})), and
total injected energy (Eq. (\ref{eq:sb_energy})) in the energy
conserving (adiabatic) phase are shown as dotted lines in (a) and (b), (e), and (f),
respectively.
Analytic predictions of SB radius (Eq. (\ref{eq:mds_radius}))
and momentum ($p_* t/\dtsn$)
in the momentum driven snowplow
phase are shown as dashed lines in (a) and (e), respectively.
Note that for $\namb$ in those equations, we use the
volume-filling WNM density, $n_w$, instead of the mean density
of the background medium, $n_{\rm avg}$,
and for $p_*$, we use $p_b(t_{\rm final})\dtsn/t_{\rm final}$, 
where $t_{\rm final}$ is the final time of each simulation.
Also note that although we do not show the analytic pressure-driven
bubble solutions, these are very close to the analytic adiabatic
solutions shown, with radius just 14\% smaller 
(see discussion in section \ref{sec:postrad}).
We also overplot as dotted lines
the predictions for 
the swept-up WNM mass ($M_{\rm sw,w}\equiv
\rho_w 4\pi r^3/3$) in (d), again using Equation (\ref{eq:sb_radius}) for
$r=r_{\rm ad}(t)$.  The circles in panel (a) denote $\tH$ and $\tHH$, the time
when $\rbub = H$ and $2H$, respectively, while the squares in panels 
(b), (c), and
(d) stand for $\tsfm$, the predicted shell formation time for a SB driven by
multiple SNe (Eq.~(\ref{eq:tsf})).  The solid horizontal lines in (g) show the
ambient medium pressure for reference.

The SBs in our simulations can be categorized by comparing two time
scales, $\dtsn$ and $\tsfm$.  The models with $\dtsn<\tsfm$ (\model{0.1}{0.01},
\model{0.1}{0.1}, \model{1}{0.01}, \model{10}{0.01}) are in
the limit of continuous energy injection, which we call the ``continuous
limit,'' while the models with $\dtsn>\tsfm$ (\model{0.1}{1}, \model{1}{1},
\model{10}{0.1}, \model{10}{1}) are in the opposite limit in which each SN acts
discretely, which we call the ``individual-SN limit.''  Model \model{1}{0.1}
does not satisfy either limit, $\dtsn \sim \tsfm$.

For the cases in the continuous limit, the overall evolution roughly
follows the analytic predictions derived in Section~\ref{sec:theory} up to
$t\sim \tsfm$ (see Figure~\ref{fig:tevol_t001}). Although the analytic
prediction assumes a uniform background medium (rather than a two-phase state),
the use of the volume-filling WNM density as the reference ambient
value ($\namb \rightarrow n_w$ for Equations
(\ref{eq:sb_radius})-(\ref{eq:tsf})) provides a good estimate for the
early-time bubble radius in most cases.  The exception is when the bubble is
big enough to enclose many cold clouds at $t\sim\tsfm$ (e.g.,
\model{0.1}{0.01}), in which case a significant amount of energy has  already
been radiated away before radiative cooling of the shocked WNM becomes
important. The bubble and hot gas masses at $t\simlt \tsfm$ are also in rough
agreement with the predicted swept-up total mass and warm gas mass,
respectively. This implies that the hot gas is mainly produced by shocks
propagating into the WNM. Although some of the shocked dense CNM clouds
undergo evaporation or ablation to supply additional mass to the interior hot
component (e.g., Model \model{0.1}{0.01}), the shocked dense clouds in
most cases
cannot remain hot because the cooling time is short at high density (see
Sections~\ref{sec:structure} and \ref{sec:pdf} for more details).  The bubble
energy is always smaller than the total injected energy even before shell
formation because of radiative losses arising from interaction of the hot gas
with dense clouds in the bubble interior. These interactions are inevitable for
SBs developing in a two-phase medium, because the forward shock
advancing through the WNM will leave dense clouds (originally CNM) behind in
the SB interior.  In flowing outward, the hot gas in the bubble interior
accelerates the dense gas with which it interacts, and loses energy by
doing work and also by mixing with dense gas (leading to radiative cooling).  
Thus, energy conserving solutions would only be strictly applicable when
the bubble expands in a single-phase (warm) medium. 

At $t\sim \tsfm$ for the models in the continuous limit, the shocked WNM
begins to cool, and the hot gas mass starts to decrease. The analytic
predictions for $\tsfm$ lie close to the time when the hot gas mass peaks in
Figure~\ref{fig:tevol_t001}(d).  After a short period of decline, the hot gas
mass again starts to rise, and the interval between SNe is short enough for
these models that the evolution remains continuous.  
In this limit, the bubble interior remains filled with hot gas (see
Figure~\ref{fig:tevol_t001}(h)) and remains at much higher pressure than the
ambient medium (see Figure~\ref{fig:tevol_t001}(g)).  The radial momentum of
the bubble continues to increase as the overpressured interior pushes the outer
shell, although the momentum increase stays far below the estimate for
non-radiative pressure-driven expansion (cf. Equations \ref{eq:shell_mom}
and \ref{eq:shell_energy}).  

In the opposite limit, the individual-SN cases (see Figure~\ref{fig:tevol_t1}),
the analytic energy-conserving continuous-injection predictions are far from
the real evolution even at early time.  Instead, the evolution due to each SN
is distinct.  The shocks propagating into both the WNM and CNM cool down, and
the bubble evolution enters the momentum conserving stage, before the next SN
explosion.  Each succesive SN heats up the bubble, and adds momentum to the
shell, but the injected energy is largely radiated away.  For Model
\model{0.1}{1} (see also \model{10}{0.1} in Figure~\ref{fig:tevol_t01}), the
remaining hot gas and the residual pressure are non-negligible so that at later
times the bubble interior remains overpressured with respect to the ambient
medium.  The bubble continues to expand and injects momentum more continuously.
However, for the extreme case of Model \model{10}{1} with very short $\tsf$,
where the bubble completely cools down before the next SNe,\footnote{ Based on
the numerical results of KO15 for the modified pressure-driven snowplow
phase, the internal pressure would drop as $P_{\rm hot}=0.8P_{\rm sf}(t/t_{\rm
sf})^{-17/7}$ after shell formation (see Equation (27) in KO15).  The pressure at
shell formation is $P_{\rm sf}=2.4\times10^6\kbol\Punit n_{\rm amb,0}^{1.26}$.
Since we assume that the heating rate is proportional to $\namb$, the pressure
of the ambient medium is $P_{\rm amb}= 1.1\times10^3\kbol\Punit n_{\rm amb,0}$.
By equating $P_{\rm hot} = P_{\rm amb}$, we obtain the time scale of ``complete
cooling'' as
$t_{\rm cool}= 22\tsf n_{\rm amb,0}^{0.11}$.
Model \model{10}{1} satisfies the condition for $t_{\rm cool}=0.3\Myr < \dtsn$,
while Models \model{10}{0.1} and \model{1}{1} have $t_{\rm cool}\sim \dtsn$.}
the pressure of the bubble is even smaller than the ambient medium so that the
bubble cannot expand further, reaching a maximum size of $\sim 130\pc$.

For the intermediate case, Model \model{1}{0.1}, the later time evolution
is similar to that of the continuous limit models, although this model
does not have a phase that is consistent with the energy conserving bubble.
Rather, the early evolution is similar to that of the individual-SN limit.

The late time evolution of the bubble radius and radial momentum is very well
decribed by the momentum driven snowplow prediction (see dashed lines in (a) and
(e)).  Although we force the coefficient to match the final momentum by using 
$p_*=\prad(t_{\rm final})/N_{\rm SN}$, the time dependences of 
$\rbub$ and $\prad$ are very close to $t^{1/2}$ and $t$, respectively.
The agreement with Equation~\ref{eq:mds_radius} is excellent for the models
in the continuous limit, but is still reasonably good in the opposite limit.

\subsection{Detailed Structure of Bubbles}\label{sec:structure}

\begin{figure}
\plotone{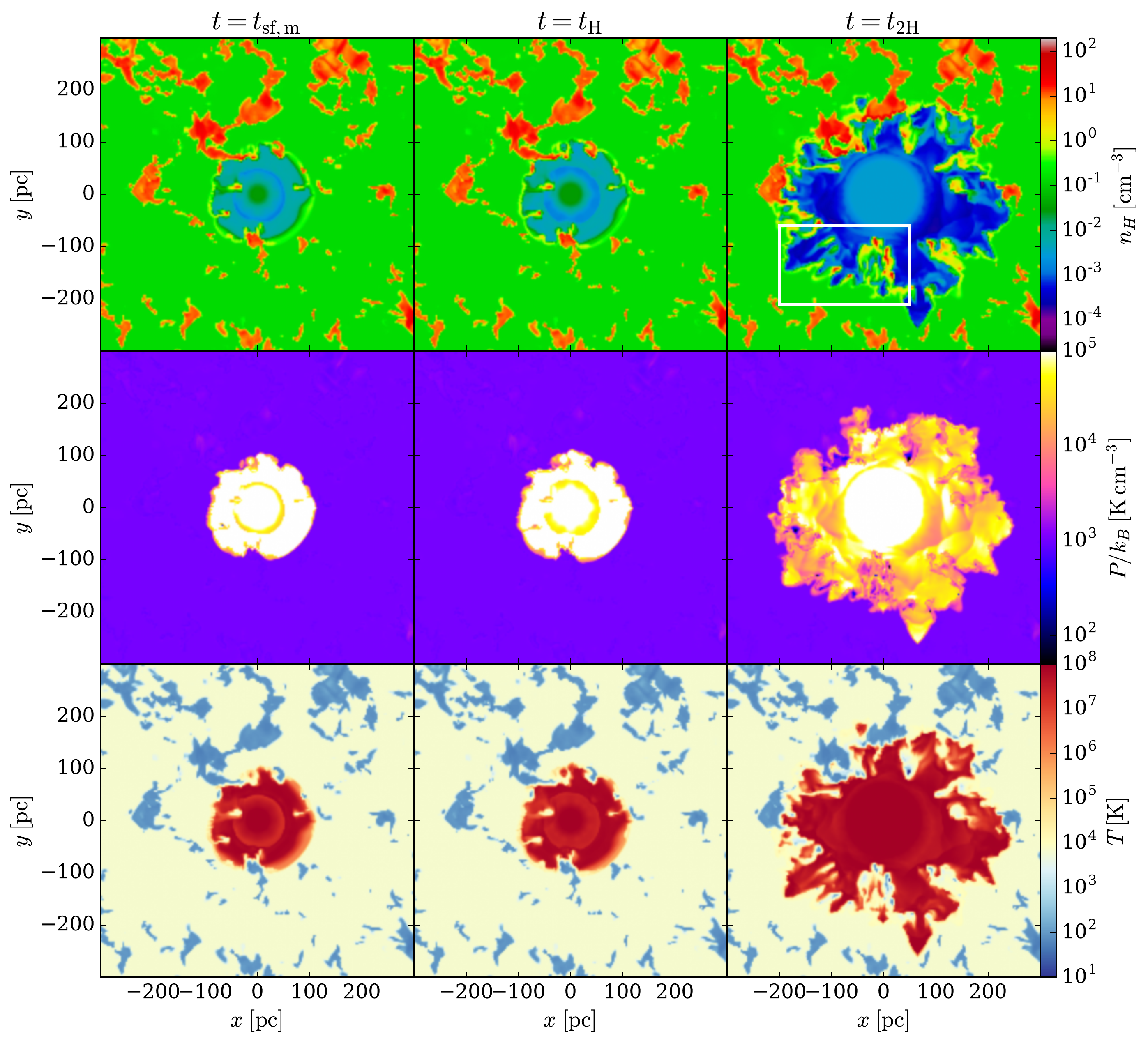}
\caption{XY-slices for Model \model{1}{0.01}. From top to bottom, logarithmic
color scales show number density, pressure, and temperature. From left to
right, columns correspond to snapshots at $t = \tsfm$, $t = \tH$, $t = \tHH$.
The white rectangle in the top-right panel indicates the region for which
zoomed images are shown in Figure~\ref{fig:zoom}.
\label{fig:slice_n1_t001}}
\end{figure}

\begin{figure}
\plotone{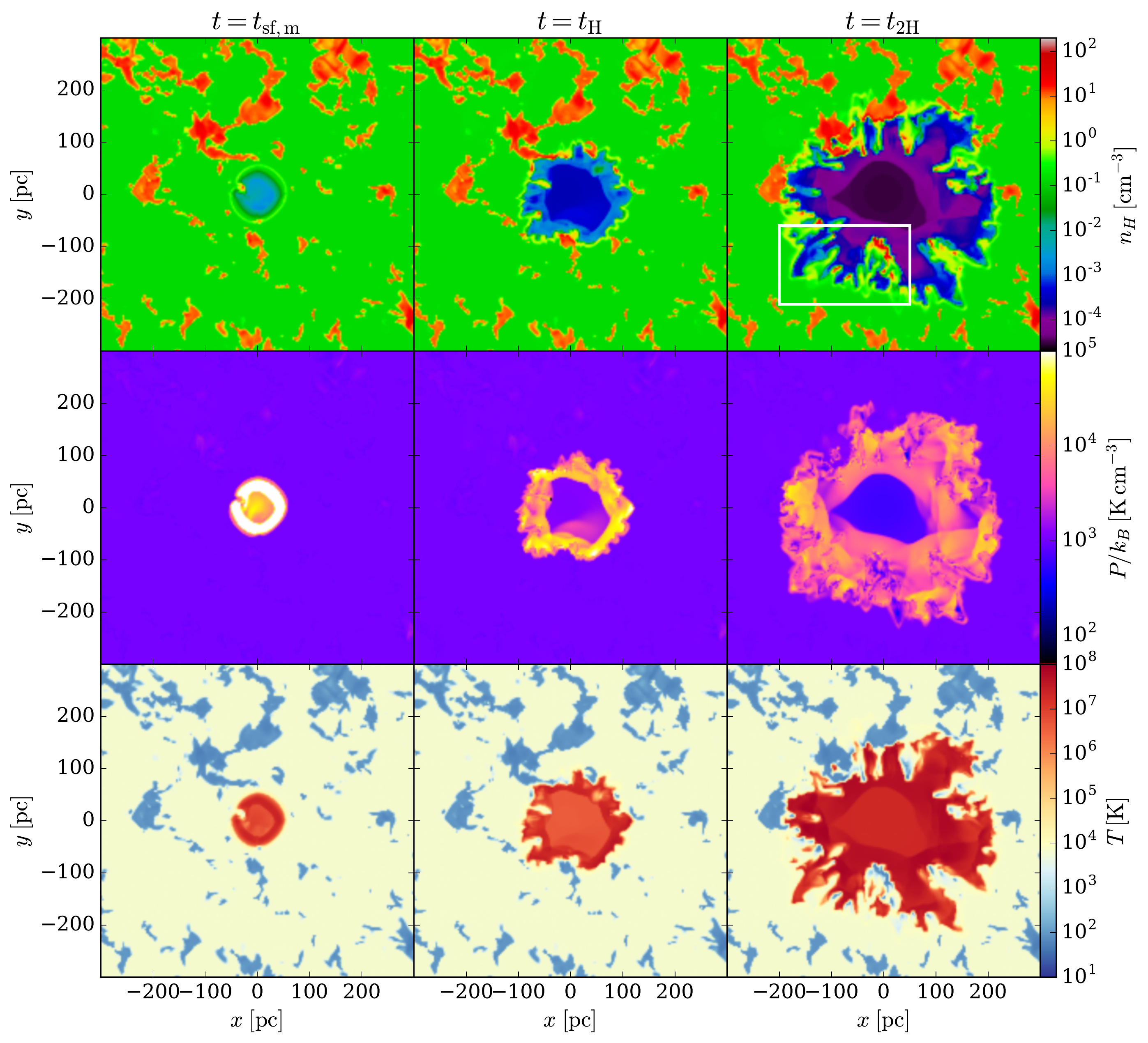}
\caption{Same as Figure~\ref{fig:slice_n1_t001}, but for Model \model{1}{0.1}.
\label{fig:slice_n1_t01}}
\end{figure}

\begin{figure}
\plotone{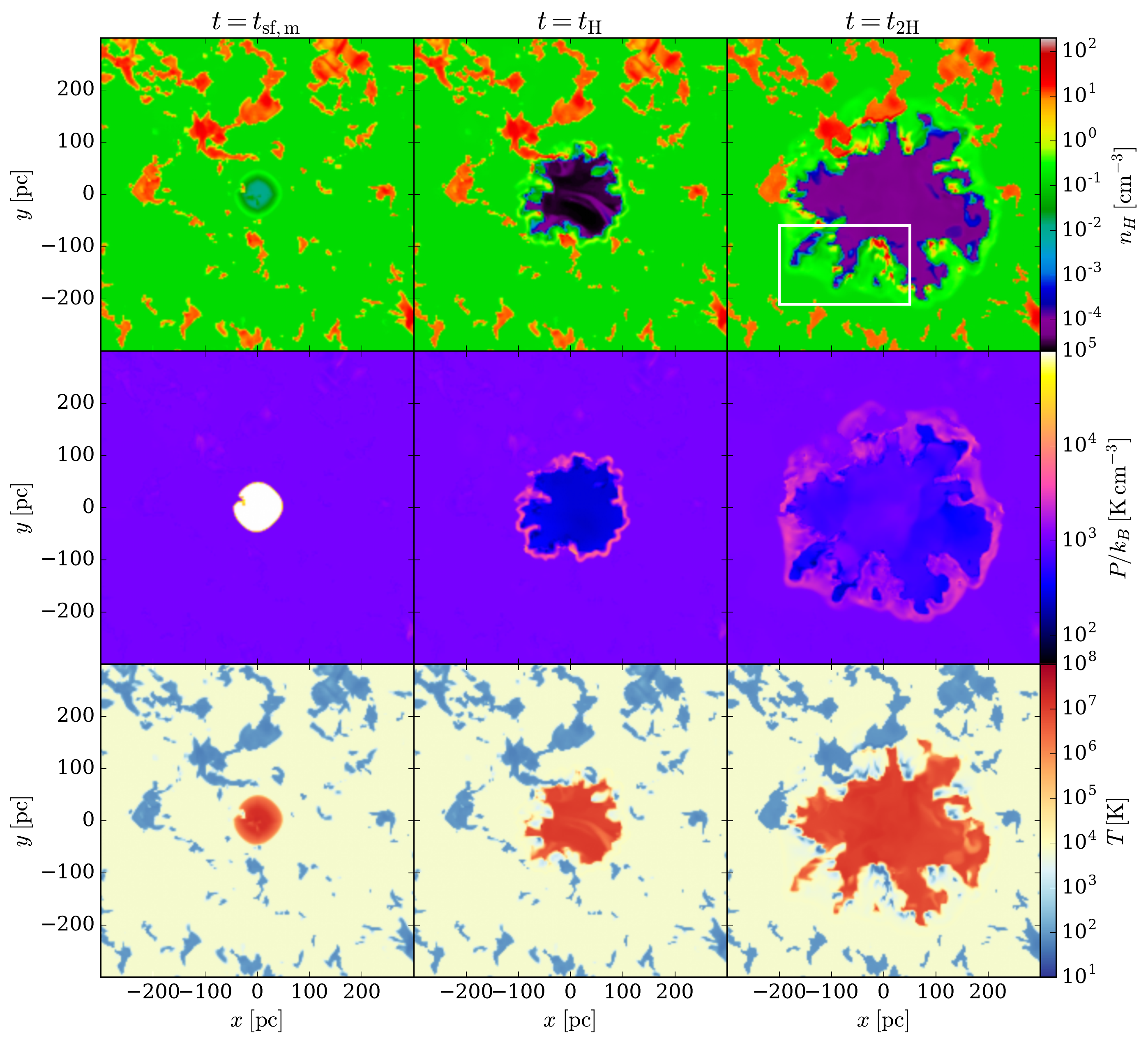}
\caption{Same as Figure~\ref{fig:slice_n1_t001}, but for Model \model{1}{1}.
\label{fig:slice_n1_t1}}
\end{figure}

To provide as sense of the evolution in SB morphology in a cloudy ambient
medium, we show slices through Models \model{1}{0.01}
(Figure~\ref{fig:slice_n1_t001}), \model{1}{0.1}
(Figure~\ref{fig:slice_n1_t01}), and \model{1}{1}
(Figure~\ref{fig:slice_n1_t1}).  Each figure consists of three rows, showing
number density, pressure, and temperature from top to bottom, and three
columns, showing snapshots at $t=\tsfm$, $t=\tH$, and $t=\tHH$ from left to
right.  We select models \model{1}{0.01} and \model{1}{1} as representative of
SBs in the continuous and individual-SN limits, respectively, while Model
\model{1}{0.1} represents an intermediate case between these limits.

Until $\tsfm$, the interior pressure is high enough that the expansion is
nearly spherical in all cases.  Since shocked WNM starts to cool earlier
when the SN rate is lower, the size of bubbles is different at $\tsfm$.

Interesting differences in morphology
can be seen in the snapshots at $\tH$ (middle columns of
Figures~\ref{fig:slice_n1_t001}-\ref{fig:slice_n1_t1}), in which 
the bubbles have similar physical sizes, but are at different evolutionary
stages. Since $\tsfm\sim\tH\sim0.3\Myr$ for Model \model{1}{0.01},
the bubble expands up to $\rbub=H$ without suffering catastrophic
energy loss. From Table~\ref{tbl:perSN} for Model \model{1}{0.01},  
44\% and 15\% of the energy that has been injected remains as total energy in
the bubble and thermal energy in the hot medium, respectively, at this time.
With $t\sim \tsfm$, the SB has retained a spherical shape and hot, highly
overpressured interior. In contrast to the case of a bubble expanding in a
uniform medium, however, there is non-negligible radiative energy loss through
shocked CNM clouds in the SB interior,
which are still dense but warm ($T\sim 10^4\Kel$). 

In contrast, for Model \model{1}{1}, the shell formed at early time
($\tsf=0.13\Myr$), and there was only one more SN event before
$\tH\sim1.9\Myr$ for this case.  Although Figure~\ref{fig:tevol_t1}(g) shows
the bubble pressure remains higher than in the ambient medium, the interior
pressure is in fact lower than in the ambient medium since the
bubble pressure is
dominated by the shell (see pressure at $t=\tH$ in
Figure~\ref{fig:slice_n1_t1}). Therefore, the shell expands in a nearly
force-free fashion (the RHS in Eq. (\ref{eq:shell_mom}) is negligible).
Radiative thin-shell instabilities
\citep{1983ApJ...274..152V,1994ApJ...428..186V} produce wiggles in the shell.
Model \model{1}{0.1} also forms a shell ($\tsf\sim\tsfm\sim0.15\Myr$) well before
$\tH\sim1\Myr$, but there were ten more SN explosions prior to $\tH\sim1\Myr$
so that the bubble interior is still overpressured and hot.  

The overall morphology of bubbles at $\tHH$ looks more or less similar in all
models, since this epoch is much later than the shell formation time
($\tHH\simgt5\tsfm$ even for Model \model{1}{0.01}). However, the detailed
internal structure and mass, momentum, and energy budgets are substantially
different. Most importantly, the bubbles still have overpressured interiors 
for Models \model{1}{0.01} and \model{1}{0.1}, while Model \model{1}{1}
has a completely exhausted interior and an overpressured shell.

\begin{figure}
\plotone{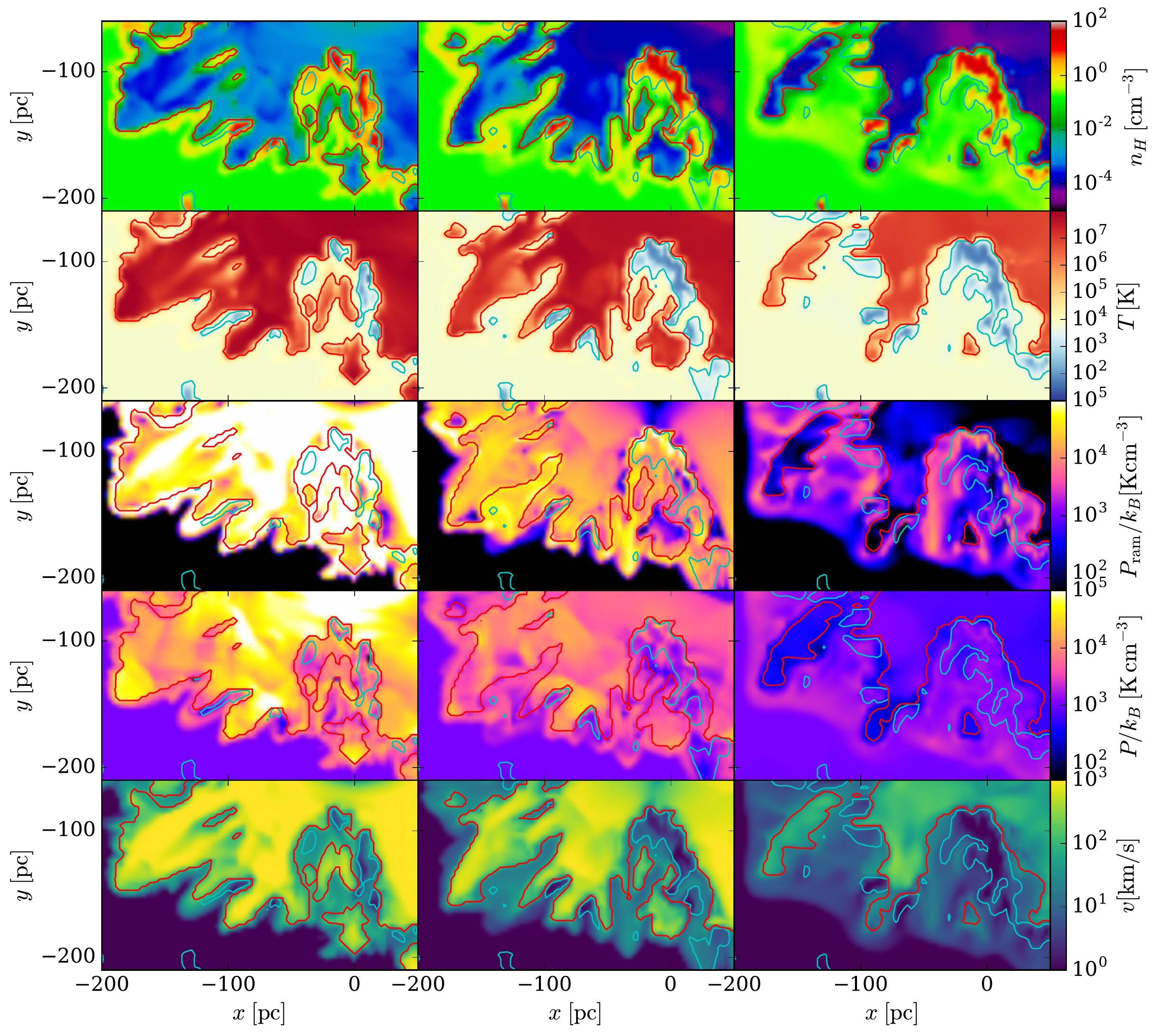}
\caption{Zoomed-in region of a patch shown in Figures~\ref{fig:slice_n1_t001}
(left column), \ref{fig:slice_n1_t01} (middle column), and
\ref{fig:slice_n1_t1} (right column).  From top to bottom, we show density,
temperature, ram pressure $P_{\rm ram}\equiv\rho v^2$, thermal pressure, and
velocity magnitude $v\equiv|\vel|$ at $\tHH$. In the panels of ram pressure,
thermal pressure, and velocity magnitude, we overplot contours of $T=500\Kel$
and $10^5\Kel$ in cyan and red that indicate cold/warm and warm/hot interfaces.
\label{fig:zoom}}
\end{figure}

To show the detailed structure and interaction between ambient medium and shell
gas (cooled bubble gas) and between shell and hot gas, Figure~\ref{fig:zoom}
displays from top to bottom zoom-in images of number density, temperature, ram
pressure $P_{\rm ram}\equiv\rho v^2$, thermal pressure, and velocity magnitude
$v\equiv|\vel|$ at $\tHH$ for the regions marked in
Figures~\ref{fig:slice_n1_t001}-\ref{fig:slice_n1_t1} (columns from left to
right).  
We also overplot isotemperature contours of $T=500\Kel$ and $10^5\Kel$ in
cyan and red to show the separation of the cold, warm, and hot phases.

The boundary between the ambient medium and the bubble is clear from
the transition in the velocity magnitude maps, while the red contours
delimit the boundary between the cooled gas in the bubble envelope and
hot interior gas.  For Model \model{1}{0.01} (left), a strong forward
shock is propagating into the ambient medium, and the interior remains
hot and highly overpressured. The bubble is bounded by a very thin
overdense shell of cooled gas.  However, for Model \model{1}{1}
(right), the thermal and ram pressure of the shocked and cooled
ambient gas exceeds that of the bubble interior, and the bubble
envelope is a broad overpressured region, rather than a thin shell.
Rather than a forward shock between the shell and ambient gas seen in
Model \model{1}{0.01}, there is smooth pressure wave propagating
into the ambient medium.

In Model  \model{1}{0.01}, there are 
embedded dense clouds that are completely surrounded
by hot gas, and some dense clouds remain warm. In Model \model{1}{1}
most dense clouds have cooled back to the cold temperature.
Model \model{1}{0.1} (middle) is intermediate, 
showing characteristics of both Models \model{1}{0.01} and \model{1}{1}.
Differences in the envelope structure (thin vs. broad shell)
are also quite clear in the top rows of 
Figures~\ref{fig:slice_n1_t001}-\ref{fig:slice_n1_t1}

We note that the evolution of dense (initially cold) clouds within SBs are
not fully resolved in the present simulations.
In our simulations, dense clouds are initially
shock-heated and accelerated when
they are overrun by the outer forward shock of the SB.  In cases with 
high-cadence SNe, these dense clouds in the interior can remain warm
due to frequent shocks from subsequent explosions, and the high
pressure of surrounding hot gas.  In cases with low-cadence SNe,
embedded clouds cool down. With extremely high resolution simulations
focused on individual clouds, 
hydrodynamical instabilities caused by shock-cloud interactions can
be followed in detail
\citep[e.g.,][]{1994ApJ...420..213K,1994ApJ...433..757M,2010MNRAS.405.1634S};
over time, these ablate small clouds and mix their material into the
bubble interior.  Here, the resolution is much more limited, and
we also neglect the thermal conduction and magnetic fields that would
affect development of instabilities that tend to destroy clouds.  
Thus, although it is uncertain exactly how limited resolution and physics 
affects the evolution of individual dense clouds in our simulations, 
we believe that our main results for the overall evolution of SBs 
are not strongly sensitive to this uncertainty.  In particular, we measure in
Appendix~\ref{sec:convergence} the hot gas mass, momentum, and energy produced
per SN at varying numerical resolution, and find these quantities are very
well converged. 

\subsection{Gas Distributions in Temperature, Velocity, and Density}
\label{sec:pdf}

\begin{figure}
\plotone{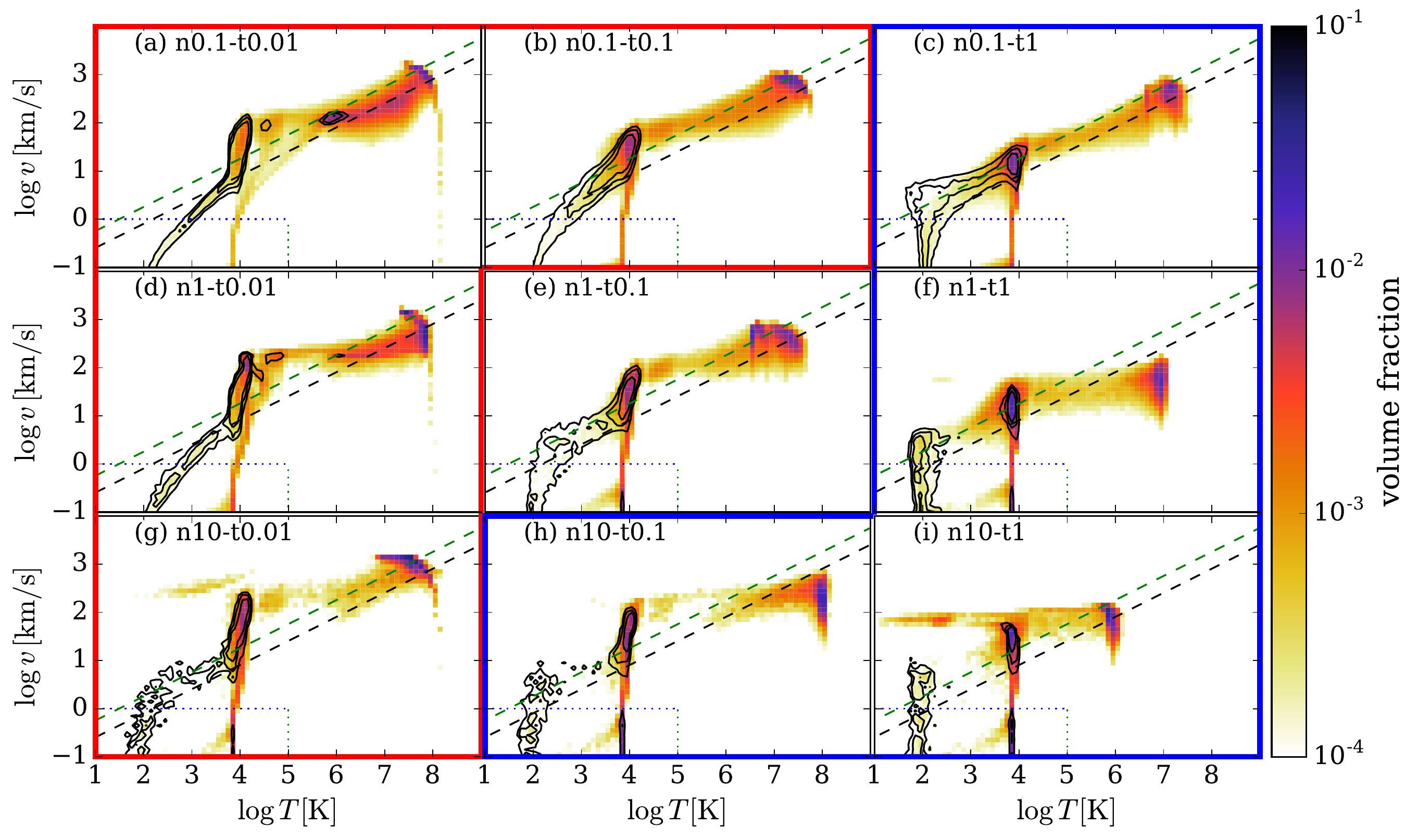}
\caption{Volume (colorbar) and mass 
(contours, $10^{-3}$, $2\times10^{-3}$, $4\times10^{-3}$ from outside to inside) 
fractions of gas 
in each bin of $\log T$-$\log v$ plane, for all
models at $t=\tH$.  The red and blue borders denote Models in continuous
($\dtsn<\tsfm$) and individual-SN ($\dtsn>\tsfm$) limits, respectively, while
Model \model{1}{0.1} in the center is intermediate ($\dtsn\sim\tsfm$).  Black
and green dashed lines denote loci of $c_s=v$ and $c_s=v/\sqrt{5}$,
respectively.  \label{fig:Tvpdf}}
\end{figure}

\begin{figure}
\plotone{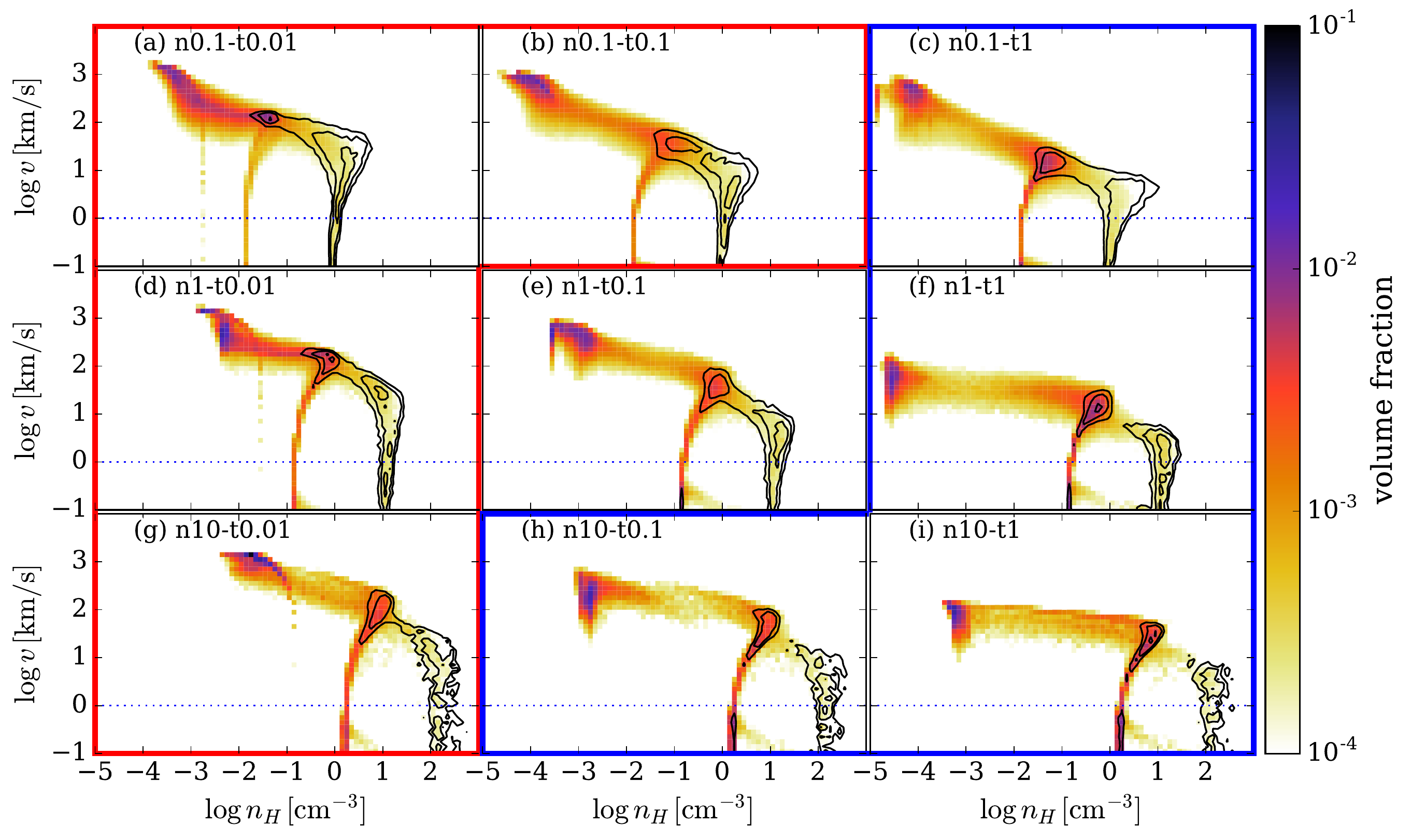}
\caption{Same as Figure~\ref{fig:Tvpdf}, but in $\log n_H$-$\log v$ plane. \label{fig:nvpdf}}
\end{figure}

We next investigate the distributions of gas in temperature, 
velocity, and density
at $t=\tH$ (i.e. when $\rbub=H$).  The probability density functions
(PDFs) provide a detailed picture of the gas that would be available to create
high speed winds when the bubble breaks out of the ISM disk into circumgalactic
space.  Figures~\ref{fig:Tvpdf} and \ref{fig:nvpdf} display the
mass (contours) and volume (colors) fractions of all the gas within $r<1.1H$
in the  $\log T$-$\log v$ and $\log n_H$-$\log v$ planes, respectively.  
In these figures, results for models that are in the continuous energy
injection limit (high SN cadence, with $\dtsn < \tsfm$)
have red borders (panels (a), (b), (d), and (g)), while results for models that
are in the individual-SN limit have blue borders (panels (c), (f), (h), and
(i)).

In Figure~\ref{fig:Tvpdf},
the dotted lines in each panel indicate the demarcation between gas that is
defined as ``ambient'' ($T<10^5\Kel$ and $v<1\kms$) and ``bubble.''  Although a
portion of the gas in the ambient regime actually consists of dense gas clouds
that have been shocked and subsequently cooled and slowed down, this
represents at most $\sim10\%$ of the total bubble mass.  Thus, while not
perfect, our definition represents a good practical criterion for
distinguishing ambient and bubble gas. In each panel, the black dashed line
shows the locus where the velocity, $v$, equals the sound speed, $c_s\equiv
(k_B T/1.27m_H)^{1/2}$. Green dashed lines show the loci where the specific
kinetic energy, $v^2/2$, equals the specific enthalpy $h\equiv \gamma
P/[(\gamma-1)\rho]=5c_s^2/2$.  Gas above and to the left of the black line is
supersonic, and gas below and to the right of the green line has the Bernoulli
parameter dominated by the enthalpy term.

The temperature-velocity distributions further distinguish different components
of the bubble gas: hot interior, shocked warm shell gas and shocked warm clouds
(originally WNM and CNM, respectively), and accelerated cold gas (shocked and
then cooled CNM clouds).  The volume-filling interior hot gas is easily seen in
Figure~\ref{fig:Tvpdf} at $T>10^6\Kel$ and $v\sim10^3 \kms$.  Moving from the
continuous-limit (top-left panels) to the individual-SN limit (bottom-right
panels), this component gets cooler and slower. The hot medium consists of gas
that was originally WNM, and was shock heated and expanded into the SB interior
to create this very hot and diffuse phase.

In Figure~\ref{fig:nvpdf},
the shocked dense clouds (originally CNM) can be found
in a vertical band at high density, also enclosed by contours.
For models with short $\dtsn$, in the continuous limit (red borders),
the dense gas has velocities up to a few tens of $\kms$.  
Although the cooling time of the shocked CNM is short due to its high density,
clouds within the bubble are repeatedly shocked and surrounded by high
pressure interior hot gas, so that the cooling is compensated by additional
shock and compression heating for models with short $\dtsn$ 
(see also the left column of Figure~\ref{fig:zoom}). Thus, these shock
accelerated dense clouds remain warm.  For the
continuous-limit models (red borders) of Figures~\ref{fig:Tvpdf} and \ref{fig:nvpdf}
 (see contours for mass-weighted PDFs),
there is no accelerated gas ($v>1\kms$)
that has returned to cold temperatures ($T\sim 10^2\Kel$) .
However, models
in the individual-SN limit (blue borders) of Figures~\ref{fig:Tvpdf} and \ref{fig:nvpdf}
show a clear distribution of cold medium with velocity $\sim 1-10\kms$
within  contours; 
this material is
dense clouds that have been shocked and accelerated, but for which the
shock and compressional 
heating is inadequate to offset cooling.  

The broad band in Figure~\ref{fig:Tvpdf} connecting the highest-temperature gas
to gas at $T\sim 10^4\Kel$ shows the effect of radiative cooling in the shell.
Shocks at the boundary of the SB accelerate WNM gas to $v\sim 100\kms$ and heat
it to high temperature, but it cools back to $T\sim 10^4$.  This creates the
warm shell of high- and moderate-velocity gas at the edge of the SB (see
Figure~\ref{fig:zoom}).  Models in the continuous limit show, in
Figure~\ref{fig:Tvpdf}, a broad warm gas distribution
with velocity range of $1-100\kms$, which is a combination of the shocked and
accelerated WNM and CNM;  in Figure~\ref{fig:nvpdf}, these components
can be distinguished based on their density. 
In models in the individual-SN limit, the warm gas is
at somewhat lower velocity, because the hot interior is lower pressure and the
expansion into the ambient medium creates weaker shocks. 

Most of the bubble gas at warm and cold temperatures is moving supersonically,
since after it was accelerated and heated in a shock, its sound speed
dropped by radiative
cooling (see Figure~\ref{fig:Tvpdf}). However, Figure~\ref{fig:Tvpdf} shows that
the hot interior gas is at most transonic in its velocities, and generally has
specific enthalpy larger than the specific kinetic energy.

\begin{figure}
\plotone{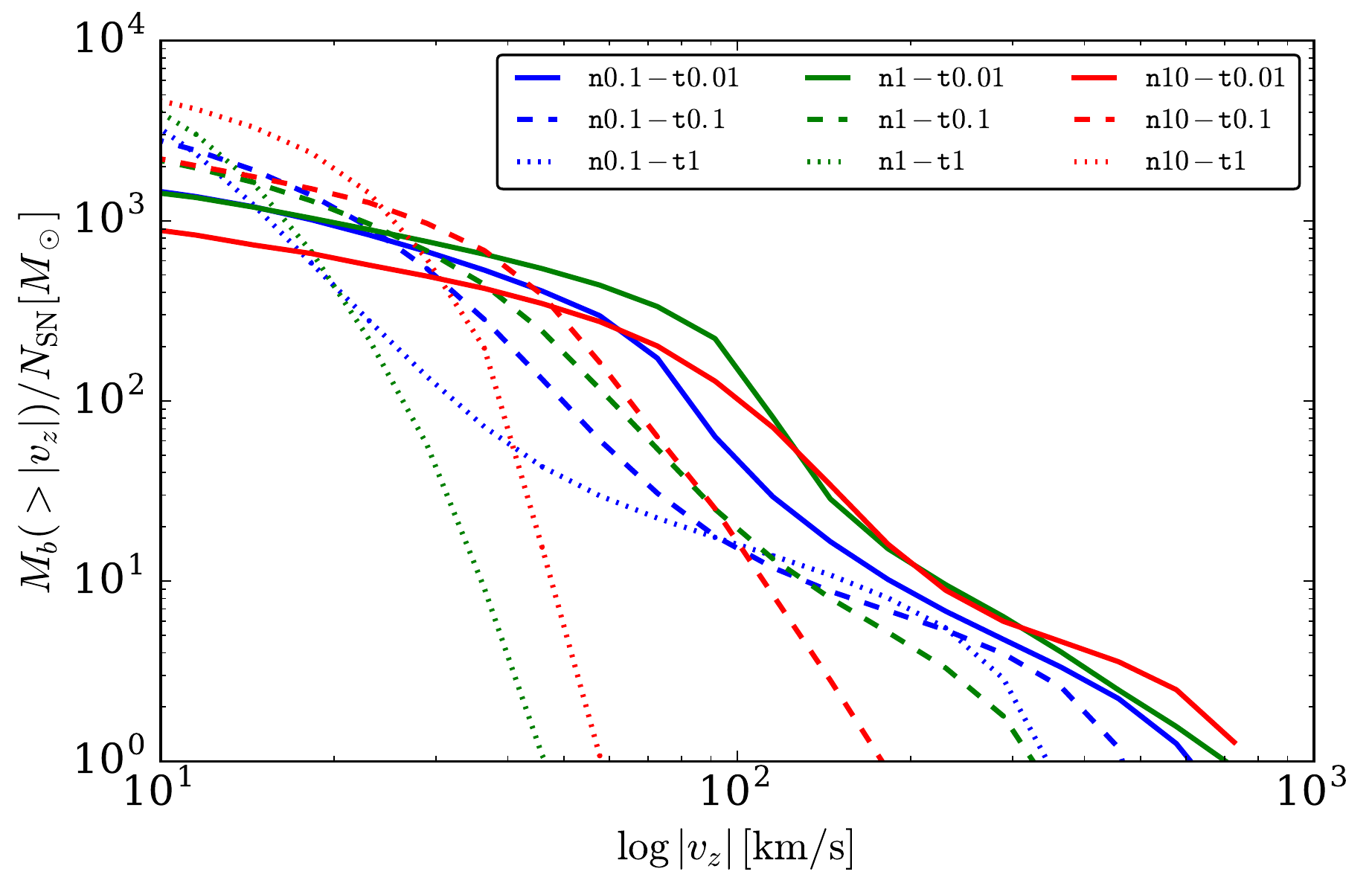}
\caption{Cumulative mass in the bubble per SN at $t=\tH$, as a function
  of velocity.  To indicate the average
  one-dimensional velocity (e.g. in the direction perpendicular to a
  disk), we use $|v_z|\equiv v/\sqrt{3}$.
\label{fig:vcum}}
\end{figure}

In addition to the mean expansion velocity of the bubble, it is also
interesting to consider the distribution of mass with velocity.
Figures~\ref{fig:Tvpdf} and \ref{fig:nvpdf} show that the velocity increases
toward lower density and higher temperature, and that the mass is divided
between the denser (and slower) former CNM and the lower density (and
faster) former WNM.  Figure~\ref{fig:vcum} plots cumulative mass (per SN)
as a function of velocity.  We use an average one-dimensional velocity 
$|v_z|\equiv v/\sqrt{3}$ to indicate e.g. the total mass that would have
vertical speed above a certain value; this is useful as an indication of
how much material could be ejected from a galactic disk.  
As is also evident in Figures~\ref{fig:Tvpdf} and
\ref{fig:nvpdf}, the velocity distribution depends more on $\dtsn$
than on $\navg$.
Except for the cases with the longest $\dtsn$, 
there is $\sim 10 \Msun$ per SN with 
$|v_z|>100-200\kms$.
As SBs are dominated by the more slowly-moving warm and cold gas, the
mass rises at lower velocity.  For the $\dtsn=0.1 \Myr$ models,
there is $\sim 100 \Msun$ per SN with 
$|v_z|> 50-70\kms$, and
for the $\dtsn=0.01 \Myr$ models, there are 
$>100\Msun$ and $> 500 \Msun$ per SN at
$|v_z|> 100\kms$ and $>50\kms$, respectively.
The gas at $|v_z|\sim 50-70\kms$ would form a galactic
fountain in a massive galaxy like the Milky Way.  However,
these results suggest that in dwarf
galaxies with shallower potential wells, 
substantial mass could escape as warm outflows driven by SBs.

\subsection{Hot Gas Mass, Energy, and Momentum Injection per SN}\label{sec:perSN}

SBs created by young, massive star clusters are one of the most plausible
drivers of galactic winds. Thus, the mass and energy budgets of SBs are of
great interest. As we have shown in Figure~\ref{fig:Tvpdf}
(see also Figure~\ref{fig:vcum}), only hot gas has
high enough velocity (higher than a few hundred km/s) that it would be able to
escape from a galaxy similar to the Milky Way.  Warm and cold gas with $z$
velocities of several tens to
a few hundred $\kms$ could, however, create a galactic fountain, while lower
velocity warm and cold gas would interact with the surrounding ISM to drive
turbulence.  In a low mass galaxy with a shallow potential, warm and cold
gas at $|v_z|\sim 50-100 \kms$ would be able to escape as a wind.

In the classical adiabatic wind model of \citet{1985Natur.317...44C}, gas is
accelerated to transonic velocities within a source region of a galaxy, and
further accelerated to escape speeds by pressure gradients as the gas expands
into circumgalactic space.  In \citet{1985Natur.317...44C} and subsequent
models of thermal-pressure-driven winds, while the combined effects of multiple
SNe are assumed to be responsible for producing the hot gas that feeds the
outflow, this is not treated directly but parameterized in terms of the mass
and energy injection per star formed (or per SN).  For adiabatic steady winds,
the conserved quantities beyond the source region are the mass flux, Bernoulli
parameter, and specific entropy.  Wind acceleration is associated with the
increase of specific kinetic energy at the expense of decreasing specific
enthalpy, while the sum of these terms (plus the gravitational potential
energy) is equal to a fixed Bernoulli parameter.  

\begin{figure}
\plotone{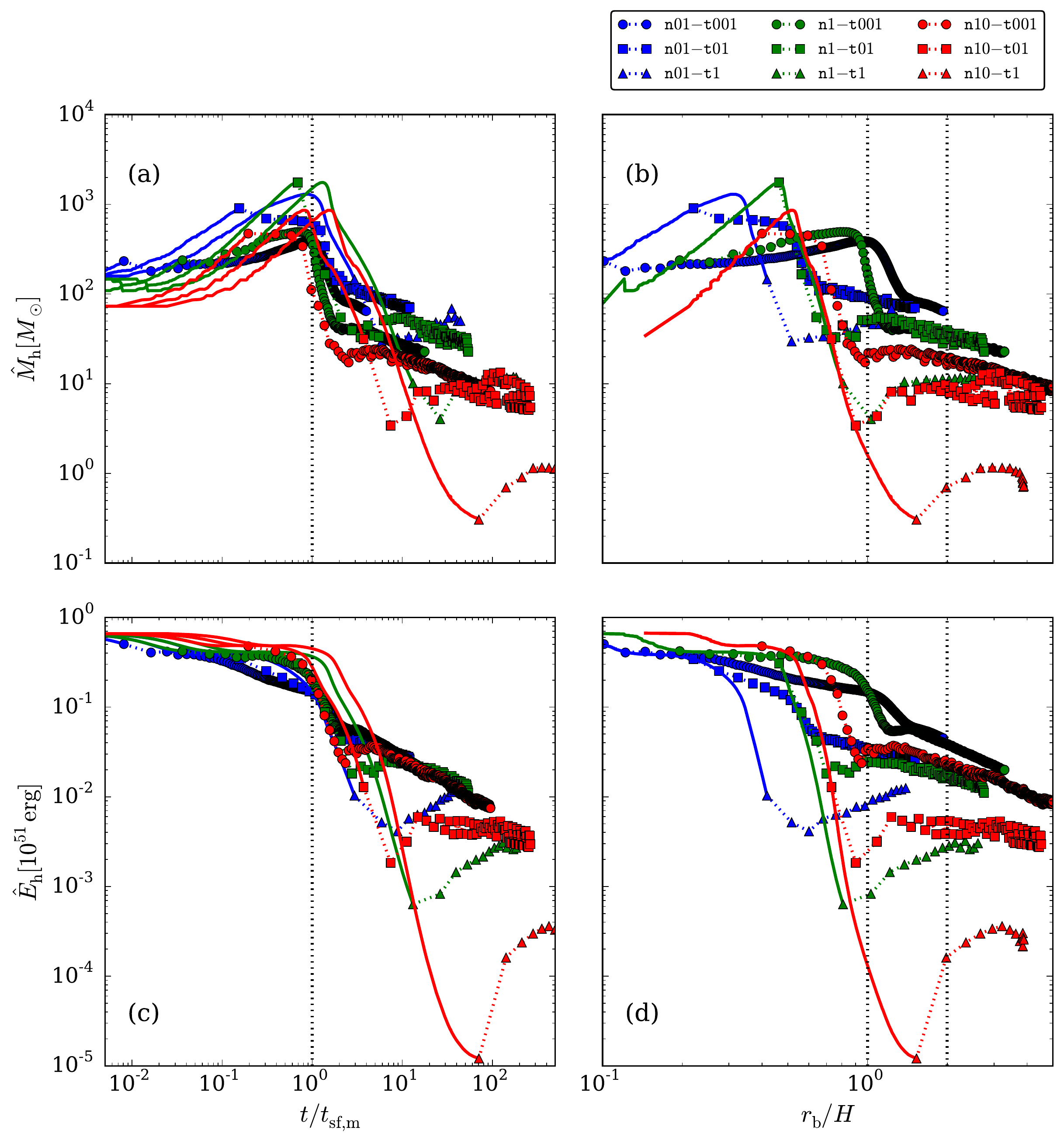}
\caption{Top: Mass of the hot gas per SN event, $\Mhat\equiv \Mhot/N_{\rm
SN}$, as a function of normalized (a) time $t/\tsfm$ and (b) radius of bubble
$\rbub/H$.  Bottom: Thermal energy of the hot gas per SN event, 
$\Ehat\equiv E_{\rm th,h}/N_{\rm SN}$ as a function of normalized (c) time
$t/\tsfm$ and (d) radius of bubble $\rbub/H$.  Blue, green, and red colors
denote the models with $\navg=0.1\pcc$, $1\pcc$, and $10\pcc$,
respectively, while circle, square, and triangle symbols denote the models with
$\dtsn=0.01\Myr$, $0.1\Myr$, and $1\Myr$, respectively.  Each symbol indicates
values at the instant immediately before each SN event.  Dotted lines connect
symbols, which provides lower/upper envelope mass/energy. We show the evolution
of the SNR from the first SN as a continuous line.  \label{fig:perSN_hot}}
\end{figure}

In Figure~\ref{fig:perSN_hot}, we plot mass ((a) and (b)) and thermal
energy ((c) and (d)) of the hot gas per SN event as functions of the normalized
time $t/\tsfm$ ((a) and (c)) and radius of bubble $\rbub/H$ ((b) and (d)).
Since evolution of bubble properties can be spiky, especially for models in the
individual-SN limit (see Figure~\ref{fig:tevol_t1}), we show as symbols only
values at the moment immediately before each SN event, and connect these
symbols with dotted lines.  The dotted lines represent lower/upper limits of
mass/energy loading.  We show the full evolution between the first and second
SNe with continuous solid lines.

As already seen in Section~\ref{sec:tevol}, the hot gas mass initially
increases rapidly as shocks propagate into the WNM, sharply drops at $t\sim
\tsfm$ when this shocked gas cools and forms a shell around the SB, and
subsequently resumes a slower increase as shocks heat the inner surface of the
shell bounding the SB and clouds left behind in the SB interior.  The evolution
of hot gas mass per SN, $\Mhat$,  reflects this
behavior.  The peaks of $\Mhat$ line up very well at $t/\tsfm\sim1$ in
Figure~\ref{fig:perSN_hot}(a), implying that Equation~(\ref{eq:tsfm}) provides
reasonably good estimates for the SB shell formation time.  The peak values
are $\Mhat\sim 500-2000\Msun$.  This is consistent with
the prediction of Equation (\ref{eq:mhsfm}) that $\hat{M}_{\rm h,sf}\sim 1000\Msun$.

Following the sharp drop
in $\Mhat$ at $t/\tsfm\sim 1$, the late stages of evolution show a slow
decline in $\Mhat$.  Except in the extreme case of Model \model{10}{1},
in which hot gas produced by each SN event completely cools down before the
next SN, the late-stage values ($t=\tH - \tHH$) of $\Mhat$ remain between
$10\Msun$ and $100\Msun$. Since we anticipate one SN for every $m_*=100\Msun$
of new stars formed from the IMF \citep[e.g.,][]{2001MNRAS.322..231K}, these
values correspond to a ``dimensionless mass loading factor''
\citep[e.g.][]{1985Natur.317...44C,2016MNRAS.455.1830T}
$\betah\equiv\dot{M}_{\rm hot}/\dot{M}_{*}=\Mhat/m_* = 0.1-1$.  Peak hot gas
mass loading values for our set of parameters are
$\betah = 5-20$, but except for cases with $\dtsn = 0.01\Myr$ and $\navg=0.1, 1$,
the time for the peak is well before $\tH$.  

SBs are expected to break out of the ISM, venting their hot gas into
circumgalactic space, when their size exceeds the scale height of the warm/cold
ISM.  Although the present simulations are for unstratified ISM disks, we can
obtain useful estimates of conditions at breakout by measuring the hot gas
properties at $\rbub=H$ and $\rbub=2H$.  These are listed in
Table~\ref{tbl:perSN} and shown in Figure~\ref{fig:perSN_hot}(b) and (d).  If
the time interval between SNe is sufficiently short (or the star cluster is
sufficiently massive), the bubble radius can reach $H$ during the energy
conserving phase, i.e. $H\le \rbub(\tsfm)$. In our simulations, Model
\model{0.1}{0.01} is only the case that satisfies this condition.  For this
model, $\betah\sim4$ at $\tH$, but $\betah$ drops to less than one before $\tHH$.
Model \model{1}{0.01} also has $\rbub(\tsfm)$ close to $H$, and has $\betah=1.7$
at $\tH$.  However, all other models have begun cooling before $\rbub$ reaches
$H$, yielding $\betah\sim 0.1-1$ at $\tH$.
For any given $\dtsn$, there is a secular decrease in $\Mhat(H)$ with
increasing density.  Similarly, for any given $\navg$, there is a 
secular decrease in $\Mhat(H)$ with increasing $\dtsn$.
However, the value of 
$\Mhat$ (and $\betah$) during breakout stages ($t\sim \tH -\tHH$)
depends more strongly on $\navg$ than on $\dtsn$. 

The dimensionless energy loading factor is defined by 
$\alphah\equiv \Ehat/E_{\rm SN}$, which is equivalent to the definition used in
\citet{2016MNRAS.455.1830T}.
In a uniform medium, by definition the total SB
energy per SN is equal to $\ESN=10^{51}\erg$ for a SB during the energy
conserving phase,
but for a multiphase ISM, some of the energy can be radiated away
even at $t<\tsfm$ via interactions with the CNM clouds.  Similarly, the thermal
energy per SN in the hot component would be fixed for $t<\tsfm$ in a uniform
medium, but not in a multiphase medium.  Figure~\ref{fig:perSN_hot}(c) shows
that $\Ehat$ declines slowly before the shell formation
time due to the cooling of shocked dense CNM clouds, and then drops
more abruptly as the shocked WNM gas begins to cool at $\sim\tsfm$.
    
\begin{figure}
\plotone{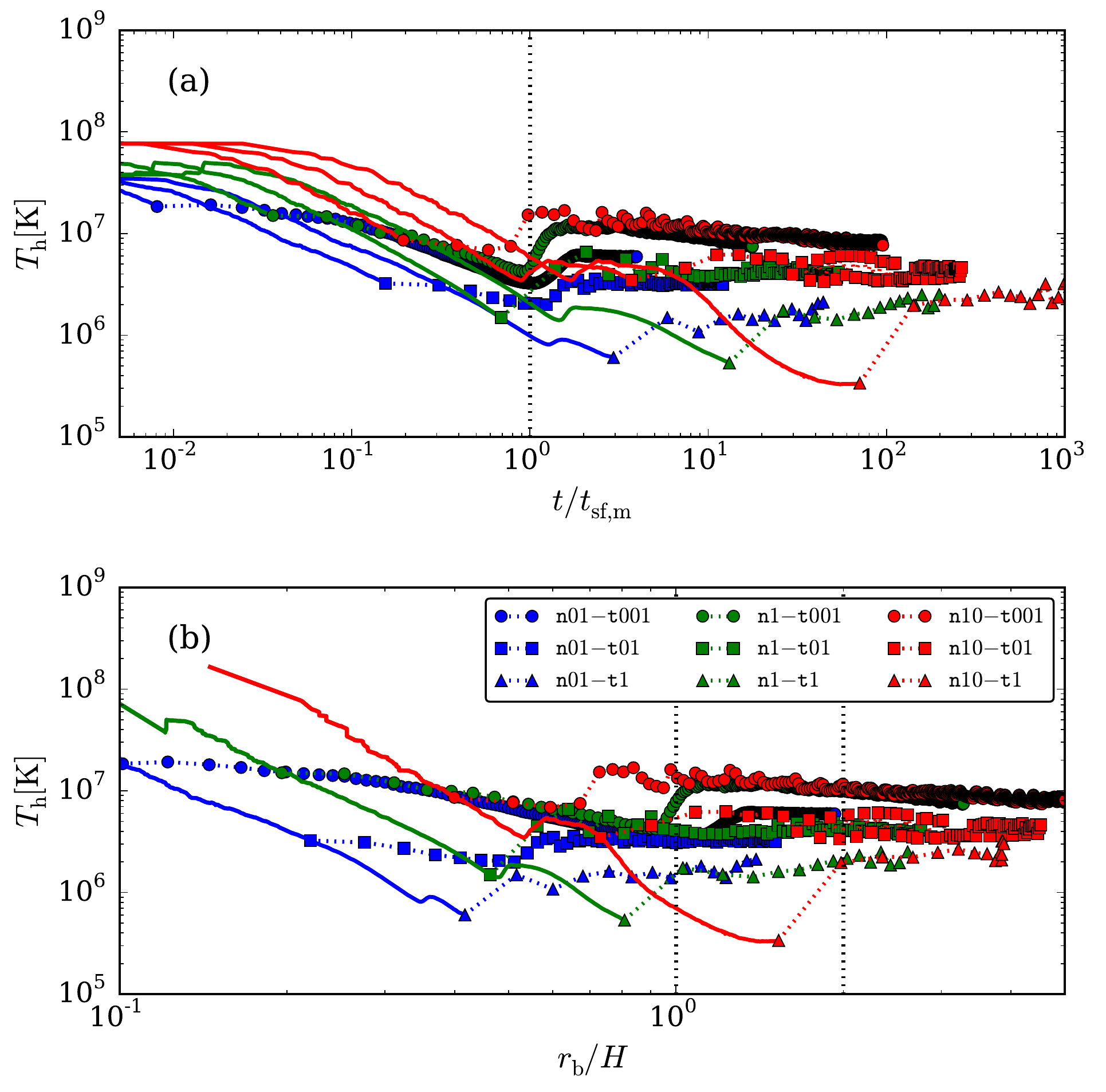}
\caption{Temperature of the hot gas immediately before each SN explosion as a
function of normalized (a) time $t/\tsfm$ and (b) bubble radius $\rbub/H$.
\label{fig:Thot}}
\end{figure}

At $\tsfm$, $\Ehat/\ESN=\alphah \sim 0.1-0.5$.
    After the strong drop in $\Ehat$ at
$t\sim\tsfm$, the subsequent decline is similar to the decline in
$\Mhat$.  In fact, after each SN event, the mean temperature of the hot
gas returns to nearly the same value (see Figure~\ref{fig:Thot}).  With nearly
constant $\Thot$, $\Ehat \propto \Mhat$.  At $\tH$, $\Ehat/\ESN$ 
has a wide range of values below $0.2$, decreasing for higher
$\navg$ and for larger $\dtsn$.  At $\tHH$, there is a narrower range of
$\Ehat$ ($\alphah \sim 0.002-0.5$ except for \model{10}{1}), and
maintains the trend of lower 
$\Ehat$ at higher $\navg$ and $\dtsn$.  

In Figure~\ref{fig:Thot}, we plot the mass-weighted mean temperature of the hot
gas, which is a key quantity for controlling large-scale wind acceleration and
escape from the galactic potential well.  For a steady flow, the Bernoulli
parameter (or function) is defined by the sum of the specific kinetic energy
$v^2/2$, gravitational potential (which is neglected here), and the specific
enthalpy $5c_s^2/2 =1.96 \kbol \Thot/\mh$ for $\gamma=5/3$ and
$\mu = 1.27 m_H$
(note that strictly speaking, $\Thot$ should be
reduced by a factor 0.4 allowing for fully ionized gas, although $c_s$ would
be unchanged).

As shown
in Figure~\ref{fig:Tvpdf}, the hot gas is mostly transonic, with enthalpy
dominating the kinetic energy in the Bernoulli parameter.  In
Figure~\ref{fig:Thot}, we only present the values of $\Thot$ immediately before
each SN event (the true evolution can be spiky as in
Figure~\ref{fig:tevol_t1}(h),
but the durations of very hot states are short).
For any given model $\Thot$ is nearly flat in the
post-shell formation stages, between $2\times 10^6 - 2\times 10^7\Kel$ for
$t=\tH - \tHH$.  
For any given $\dtsn$, the range of $\Thot$ for $t=\tH - \tHH$
is even smaller, and $\Thot$ increases with decreasing $\dtsn$.
This suggests that the enthalpy of the hot gas that loads
winds would be insensitive to exactly when and how breakout occurs.
Furthermore, $\Thot$ during the breakout stage 
depends more on the mass of the cluster driving the outflow (i.e. on
$\dtsn$)
than on the conditions of the ambient ISM ($\navg$).  Note that this behavior is
opposite to the hot gas mass loading,
which depends more strongly on $\navg$ than
on $\dtsn$ (compare Figure~\ref{fig:Thot}(b) with Figure~\ref{fig:perSN_hot}(b)).  However, Figure~\ref{fig:vcum} shows that the overall distributions of
mass with velocity are more sensitive to $\dtsn$ than $\navg$.  

In addition to loading of winds, SBs are important for driving turbulence in
the warm/cold ISM, which in turn regulates SFRs.  For self-regulated disk star
formation, the turbulent pressure is proportional to the mean
momentum injection per
unit stellar mass formed $p_*/m_*$, while the SFR is inversely proportional to
$p_*/m_*$ \citep{2011ApJ...731...41O,2011ApJ...743...25K}.
Previously, KO15 measured the final radial momentum of late-stage SNRs from
single SNe in two-phase ISM backgrounds with a large range of $\navg=0.1-100$, 
as well as a few different cases with multiple SNe and $\dtsn =
1\Myr$.  Here, we quantify momentum injection in terms of the mean radial
momentum per SN for all our models.

\begin{figure}
\plotone{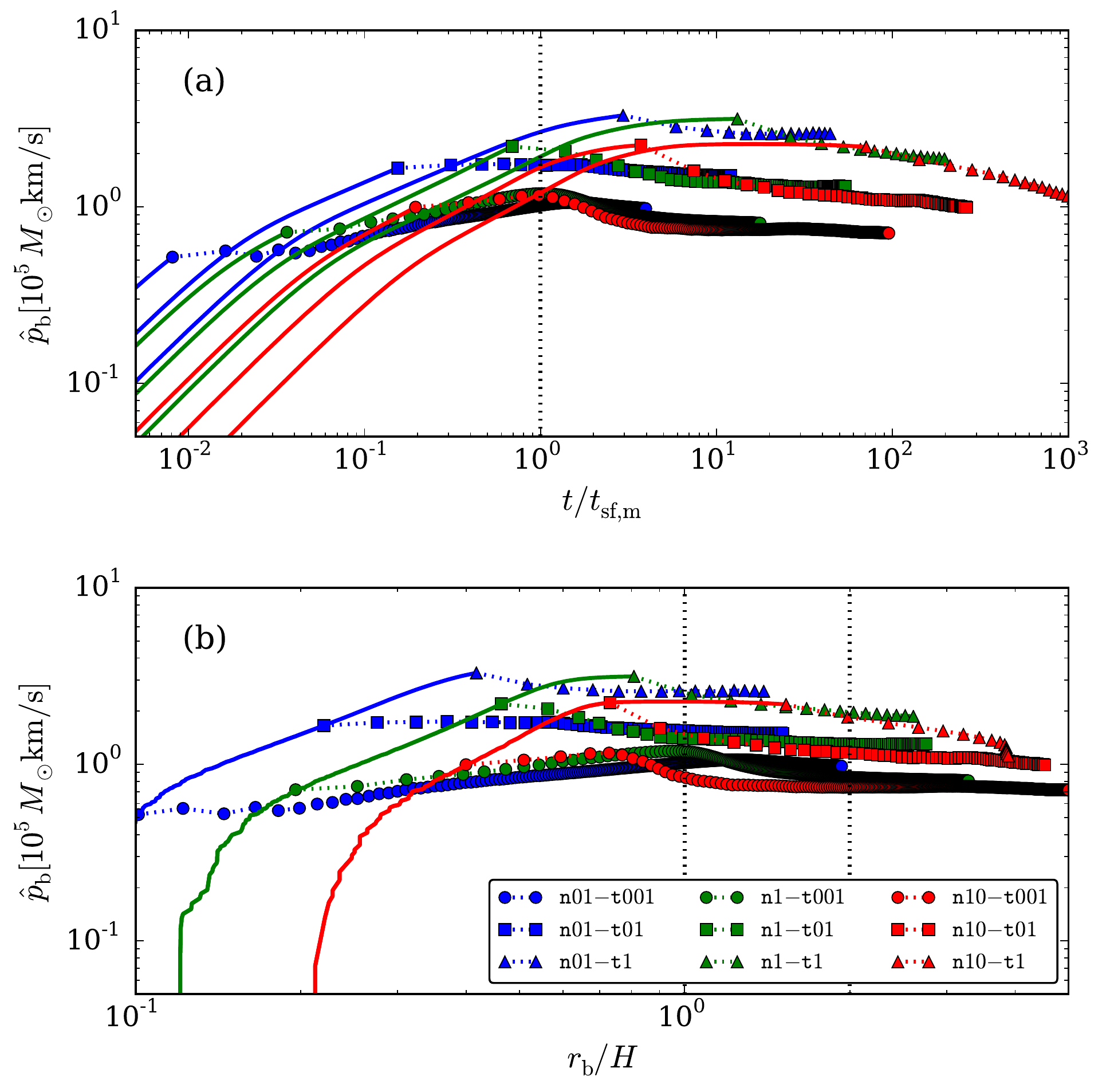}
\caption{ Total radial momentum of the bubble per SN event, $\phat\equiv
\prad/N_{\rm SN}$, as a function of normalized (a) time $t/\tsfm$ and (b)
bubble radius $\rbub/H$.  \label{fig:perSN_p}}
\end{figure}

Figure~\ref{fig:perSN_p} shows $\phat$, the radial momentum of the SB per
SN, as functions of (a) normalized time and (b) normalized radius.
At $\tsfm$, the values of $\phat$ are comparable to the
prediction of Equation (\ref{eq:psfm}).  For all
models, $\phat$ declines slightly after $\tsfm$, but generally evolves
very weakly at late stages, and is quite insensitive to parameter values.
For single
SNe, KO15 showed that the final momentum is $\sim3\times10^{5} \Msun\kms$ for
$\navg=1\pcc$, and weakly decreasing $\propto(\navg/1\pcc)^{-0.17}$. 
Here, our models with $\dtsn=1\Myr$ have similar $\phat$ 
to the single-SN results at $\tH$, while $\phat$ is lower at
$\dtsn=0.1\Myr$ ($\sim 1.5\times 10^5\Msun\kms$) and $\dtsn=0.01\Myr$ ($\sim 1
\times 10^5\Msun\kms$).  There is also a slight ($<50\%$) decrease in $\phat$ 
from $t=\tH$ to $\tHH$. The dependence of $\phat$
on $\navg$ is even weaker than in the single-SN case.

\section{IMPLICATIONS FOR LOADING OF GALACTIC WINDS}\label{sec:windload}

In Section \ref{sec:perSN}, we provided results for the mass of hot
gas per SN as a function of time and radius
(Figures~\ref{fig:perSN_hot}(a) and (b)).  Table~\ref{tbl:perSN} shows that
except for the models that have $\tH \simlt \tsfm$, $\Mhat$ is
relatively constant for $\rbub\sim H - 2H$ for any individual SB, and lies
in the range $10 - 100 \Msun $ for the parameter set considered, with
the lower end corresponding to ISM disks with larger $\navg$. As
discussed in Section \ref{sec:theory_breakout}, the expanding shells
of SBs from sufficiently massive clusters with short $\dtsn$
are likely to remain coherent until breaking out of the disk, whereas
SBs driven by lower mass clusters with long $\dtsn$ will have shells
that merge with the turbulent ISM prior to breaking out.

Even if the outer shell of a SB does not maintain its integrity, the
high-entropy hot gas in the interior will tend to rise and make its way out
of the galaxy.  Since not
all of the hot gas created in a SB will ultimately be able to escape, an upper
limit on the contribution from each SN to a hot wind is $\Mhat$.
Dividing by a typical mass of stars $m_*= 100\Msun$ formed per SN,
this implies that the hot wind ``mass loading'' factor
$\betah = \Mhat/m_*$ would be less
than unity unless the conditions of the ISM and clusters driving SBs
combine to enable the SB radius to exceed $H$ before $\tsfm$.  With
velocities of warm gas in the shell only up to $\sim 100\kms$ (see Figures
~\ref{fig:Tvpdf} and \ref{fig:nvpdf}), this warm gas could not immediately escape as a
wind from a massive galaxy, although in principle some of this
material could be further accelerated by interaction with the faster
hot gas or cosmic ray wind that is flowing out of a galaxy.  As noted
earlier, at $\tH$ the total mass of gas with $|v_z| \simgt 50 \kms$ exceeds
$100\Msun$ for the models with $\dtsn =0.1, 0.01 \Myr$, implying
that for dwarf galaxies more material (mostly at warm temperatures)
could escape as a SB-driven outflow than is locked up in stars.

Given the low $\betah$ values for our models with $\tsfm < \tH$, we
suggest that a heavily mass-loaded hot wind (i.e. $\betah>1$ in the hot
component) is only possible if conditions enable ISM breakout prior to shell
formation.
Furthermore, from Equation (\ref{eq:mhsfm}), since the maximum mass in the SB at
shell formation is $\sim 10^3\Msun$ and not all of this gas would escape,
there is an upper limit $\betah \simlt 10$ for SN-driven hot winds.
Setting $\rsfm=H$ 
and solving for $\dtsn$ (using Equation \ref{eq:rsfm}), the maximum interval
(in Myr) between SNe that is consistent with the ``hot break-out'' condition is
\begin{equation}\label{eq:dt_hbo}
\Delta t_{\rm hbo,6}=0.019 E_{\rm 51} (f_{w,-1}n_{\rm avg,0} )^{-1.7} H_{2}^{-2.7}.
\end{equation}
Here, $H_2\equiv H/100\pc$ and we use
$\namb=n_w\approx f_w\navg$ if the volume
fraction of the CNM is negligible, where
$f_w$ is the mass fraction of the WNM and $f_{w,-1}\equiv f_w/0.1$.
\footnote{If the ISM is primarily molecular, rather than two-phase
  atomic, $\namb/\navg$ could instead be computed
  based on the variance in the density PDF, and would scale inversely
  with the Mach number of the turbulence.}

In Section \ref{sec:method} we adopted Equation (\ref{eq:H_def}) for the typical
ISM scale height, but this can be generalized under the assumption of
vertical dynamical equilibrium in the ISM to 
$H=\sigma_z[\pi^2 G\rho_{\rm avg}(1+\chi)]^{-1/2}$ with $\chi$
(approximately)
the ratio of midplane stellar+dark matter density to mean midplane gas density
under typical disk conditions
\citep[][]{2011ApJ...731...41O,2015ApJ...815...67K};
$\chi \sim 1$ in the Solar neighborhood, but
gas may dominate in starburst regions.
In addition, the mean midplane density is related to the total midplane
pressure by $\rho_{\rm avg} = P_{\rm tot}/\sigma_z^2$, giving
$H=\sigma_z^2 [\pi^2 G P_{\rm tot}(1+\chi)]^{-1/2}$. 

Over long timescales, analytic theory \citep{2010ApJ...721..975O,
  2011ApJ...731...41O, 2011ApJ...743...25K} predicts, and numerical simulations
\citep{2011ApJ...743...25K,
  2013ApJ...776....1K,2015ApJ...815...67K}
have verified, that the ISM will evolve to an equilibrium
state that is self-regulated by feedback from star formation, in which
$P_{\rm tot}$ is approximately linearly proportional to
the star formation rate per unit area, $\Sigma_{\rm SFR}$.  
Based on theory and simulations, the expected total feedback yield
$\eta\equiv P_{\rm tot}/\Sigma_{\rm SFR}\approx 10^3\kms$; we define
$\eta_{3}\equiv \eta/10^3\kms$.  The normalized
density and scale height can then be written as
\begin{equation}
n_{\rm avg,0} = 0.28\, \eta_{3} \sigma_{z,1}^{-2} \Sigma_{\rm SFR, -3} 
  \end{equation}
and
\begin{equation}
H_{\rm 2} = 3.5\, \eta_{3}^{-1/2} \sigma_{z,1}^{2} \Sigma_{\rm SFR, -3}^{-1/2} 
  \end{equation}
where $\sigma_{z,1} \equiv 10\kms \sigma_{z}$, 
$\Sigma_{\rm SFR, -3} \equiv \Sigma_{\rm SFR}/10^{-3} \Msun \kpc^{-2} \yr^{-1}$ and
we set $\chi=1$ for convenience.

Assuming that the background ISM state is consistent with
self-regulated equilibrium, the limiting SN interval that allows
hot break-out can then be computed using Equation (\ref{eq:dt_hbo}), and
the corresponding minimum star cluster mass (using Equation \ref{eq:dtsn})
would be
\begin{equation}\label{eq:M_hbo}
  M_{\rm cl, hbo}= 6.6\times 10^5\Msun \, E_{\rm 51}^{-1} f_{w,-1}^{1.7} \eta_{3}^{0.35} \sigma_{z,1}^{2.0}\Sigma_{\rm SFR, -3}^{0.35}  
\end{equation}
with the corresponding SFR obtained by dividing by $t_{\rm life}=40 \Myr$.
For Solar neighborhood conditions, where $\Sigma_{\rm SFR,-3} \sim 3$, a 
very massive cluster ($\sim 10^6 \Msun$) would be required to
enable hot breakout.

In fact, the SN that drive a SB need not all originate in a single cluster.
Several clusters that are born within $t_{\rm life} \sim 40 \Myr$ of each other, at
distances $\simlt H$, effectively act like a single cluster from the
point of view of driving a SB \citep[e.g.][]{2016arXiv160300815Y}.
  It is therefore useful to compare Equation
(\ref{eq:M_hbo}) with the average
total mass of recently-formed local stars that would contribute to a single SB
(under self-regulated equilibrium, and again taking $\chi \sim 1$),
\begin{equation}\label{eq:m_young}
  \langle M_{\rm young,H}\rangle \equiv t_{\rm life} \pi H^2 \Sigma_{\rm SFR}
  =1.5\times 10^4\Msun \,  \eta_{3}^{-1} \sigma_{z,1}^{4}.
\end{equation}
For fiducial $\sigma_z$ and $\eta$,
the corresponding mean SN interval and SFR within
$\pi H^2$ are $\Delta t_{\rm SN,H}=m_*/(\pi H^2 \Sigma_{\rm SFR})\sim 0.3 \Myr$
and $\dot{M}_{*,H}= \pi H^2 \Sigma_{\rm SFR} \sim 4 \times 10^{-4}\Msun \yr^{-1}$, respectively.  Note that these are independent of the local gas surface density.

A large upward fluctuation in the local star formation rate
would be needed to increase the local mass in young stars by a factor
$\sim 40 \sigma_{z,1}^{-2} \Sigma_{\rm SFR, -3}^{0.35}$ from the typical value in 
Equation (\ref{eq:m_young}) to the level required for hot breakout by
Equation (\ref{eq:M_hbo}).  Although the required level of upward fluctuation
is higher in regions of increased $\Sigma_{\rm SFR}$, this may be partly
compensated if $\sigma_z$ also increases  under these conditions.
Indeed, while in observed disk galaxies
$\Sigma_{\rm   SFR}$ varies by several orders of magnitudes and $\sigma_z$ varies
by only a factor of a few, the variations are observed to be
correlated
\citep[e.g.][]{2009AJ....137.4424T,2011MNRAS.410.1409W,2013ApJ...773...88S,2015AJ....150...47I}.  Nevertheless, unless most of the star formation
in galaxies occurs in bursts that are well above the time-averaged star formation
rate, SBs will generally undergo shell formation before  breakout and 
the SN-driven hot winds they create will only have a mass loading factor
$\betah \sim 0.1 - 1$. 

Starburst galaxies have very high central concentrations of gas, and
correspondingly
quite high localized values of $\Sigma_{\rm SFR}$.  Although these conditions
are much more extreme than typical regions in galactic disks, the relationship
between ISM equilibrium pressure (or weight) and the mean value of
$\Sigma_{\rm SFR}$ still appears to be consistent with the prediction of
self-regulation by SN feedback
\citep{2011ApJ...731...41O,2012ApJ...754....2S,2012MNRAS.421.3127N}.  
Equation (\ref{eq:M_hbo}) would therefore still represent the minimum mass of
young stars within $\sim \pi H^2$ that is needed for a  burst to
produce a hot breakout. For starburst regions
with $\Sigma_{\rm SFR,-3}=10^2-10^5$, this corresponds to 
$M_{\rm cl, hbo} \sim 3\times 10^6- 4\times 10^7\Msun$
or $\dot{M}_{\rm *, hbo}\sim 0.1-1 \Msun \yr^{-1}$.  While
assessment of the observed scale height or velocity dispersion of the
atomic/molecular ISM in galactic centers
is challenging due to limited resolution
\citep[but see][]{2015ApJ...801...25L},
observed galaxies with winds powered by central starbursts
do have total $\dot{M}_* \sim 0.1-10^2 \Msun \yr^{-1}$ within 
the central few hundred pc \citep{2015ApJ...809..147H}. Intriguingly,
the observed values of $\beta$ in these starburst-driven winds decrease with
increasing SFR, perhaps reflecting the greater difficulty of
achieving hot breakout under the higher-density conditions that yield higher
$\Sigma_{\rm SFR}$ (as evident in the increase of $M_{\rm cl,hbo}$ with
$\Sigma_{\rm SFR}$ in Equation \ref{eq:M_hbo}).

We conclude that the equilibrium SFR, based on a temporal and spatial averages,
is in general too low to drive a heavily loaded hot wind.  Nevertheless,
a massive cluster or large-amplitude fluctuation in $\Sigma_{\rm SFR}$ could
in principle lead to a hot outburst with maximum $\betah \sim 10$, and this
appears to occur in nuclear regions for starburst-driven outflows.
More typically, we expect $\betah \sim 0.1 -1$
for SN-driven hot winds on large scales in disk galaxies.
For disk-launched winds, the mass-loss
rate per unit area on each disk face would be $\betah \Sigma_{\rm SFR}/2$,
whereas for quasi-spherical
nuclear winds the total mass-loss rate would be $\betah \times \dot{M}_*$.  

Finally, we note that for SN-driven steady-state hot winds,
the flow velocity at large distance is obtained from the Bernoulli
parameter
${\cal B} \equiv (1/2) v^2 + (5/2)P/\rho  + \Phi$, which is
constant along streamlines for an adiabatic flow.  For the hot gas within
SBs, the enthalpy term dominates (see Figure~\ref{fig:Tvpdf}).
However, after breakout, as
streamlines expand and $P/\rho $ decreases ($\propto (v r^2)^{-2/3}$ for a
spherical flow), the flow will accelerate and the
kinetic term will begin to dominate.
Neglecting the potential term, at large distance the velocity would
approach $v_{\rm asy} = (2 {\cal B})^{1/2}$, where ${\cal B}$ is set by
the enthalpy of hot gas in the SB interior prior to breakout.
For the range of values of $\Thot(H)$ and $\Thot(2H)$ in Table~\ref{tbl:result},
$v_{\rm asy} =(5 P/\rho)^{1/2} = (3.9 k_B\Thot/m_H)^{1/2}$ is in the
range $200 - 600 \kms$. This implies that SB-driven hot winds can
escape at high velocity from the immediate vicinity of all but the most
massive galaxies.

For SBs at $t > \tsfm$, the effective momentum per
unit time that the successive
SNe impart to their surroundings is equal to $\phat/m_*$ multiplied by 
the SFR.  From the results for $\phat$ in Table~\ref{tbl:perSN},
and using $m_*=100 \Msun$,
this is $(1-2) \times 10^3 \kms$ multiplied by the SFR. If this
momentum is equally
shared with all of the surrounding gas within the disk scale height,
the mean velocity at breakout will be comparable to the turbulent velocity
dispersion in the disk -- at most several tens of $\kms$ (see Equation
\ref{eq:vmds_H} and following, and the values for $v_b$ in
Table~\ref{tbl:result}).
However, the initial breakout of a SB can clear
much of the surrounding ISM.  The time required for initial breakout, using
the results of Section \ref{sec:theory_breakout}, is
$(H/\sigma_z)(\dtsn/6 \Delta t_{\rm SN, H})^{1/2}$.  For regions
where the dynamical time $H/\sigma_z$ is shorter than $t_{\rm life}$,
energy and momentum input from SNe will continue, but the
momentum flux in the vertical direction will be shared with much less
material.  In this situation, a low value of $n_{\rm amb,0}$  in Equation
(\ref{eq:mds_velocity}) can lead to a very fast outflow.

\section{SUMMARY}\label{sec:sumndis}

The energy released by SNe is vital to the ISM and to the surrounding
CGM and IGM on larger scales, and understanding the interaction of
clustered SNe (the typical case) with their environment is essential
to theories of both the ISM and galaxy formation.  In this paper, we
have used numerical simulations to study the evolution of SBs driven
by multiple SNe as they expand into the two-phase (warm/cold) ISM,
which in our simulations has realistic complex cloudy structure that
results from saturation of thermal instability.  We consider models
with a range of mean background density $\navg=0.1 - 10 \cm^{-3}$, and
interval between SNe $\dtsn = 0.01 - 1 \Myr$.  The former corresponds
to a typical range of gas surface density $\Sigma_{\rm gas} \sim 5 -
50 \Surf$ and star formation rate surface density $\Sigma_{\rm SFR}
\sim 4 \times 10^{-4} - 4 \times 10^{-2} \Msun \kpc^{-2} \yr^{-1}$.
The latter corresponds to a range of star cluster mass (or total local
mass in young stars) of $\Mcl \sim 4 \times 10^3 - 4 \times 10^5
\Msun$.  Our simulations are idealized in that we do not include
background stratification of the mean density and pressure.
However, we can use expected relationships between mean midplane
density and ISM scale height $H$ to define the times $\tH$ and $\tHH$
when the SB radius reaches $H$ or $2H$, such that if stratification
were included the SB would break out of the warm/cold disk into the hot
corona.  We
measure key SB properties -- total radial momentum of the bubble $\prad$,
hot gas mass $\Mhot$, and hot gas temperature $\Thot$ -- at
times up to $\tHH$.  Taking ratios with the total number of SN events
that have occurred, we compute $\phat$ and $\Mhat$, the
momentum and mass of hot gas injected per SN; we tabulate these at
$\tH$ and $\tHH$ as $\phat(H)$, $\phat(2H)$, etc. (see
Table~\ref{tbl:perSN}).

Our main conclusions are as follows:

\begin{enumerate}

\item {\it Evolution}

  As in the case of a SNR from a single SN, a
  blast driven by multiple SNe initially evolves similarly to analytic
  predictions for adiabatic expansion.  Equation (\ref{eq:tsfm})
  provides a prediction for the time $\tsfm$ when a cooled shell will
  form at the leading edge of the blast wave; this assumes continuous
  energy ejection, with $\dtsn < \tsfm$.
  Figures~\ref{fig:tevol_t001} - \ref{fig:tevol_t1} show that the mass
  in hot gas peaks at $t \sim \tsfm$ for models with short $\dtsn$.
  After shell formation, SB radii expand more slowly than the
  classical prediction for an adiabatic pressure-driven snowplow.  This is
  because energy is lost from the hot interior through cooling (due to
  mixing with material ablated from embedded dense clouds, and at the
  irregular interface with the cooled shell).  For models with
  $\dtsn=1\Myr$, evolution behaves like a succession of individual
  events (with strong cooling after each one), whereas the evolution
  is continuous in models with $\dtsn=0.01\Myr$.  For our set of
  parameters, the SB radius expands to $H$ within $\sim 1-10\Myr$
  (see Table \ref{tbl:result} for the values of $\tH$ and $\tHH$).
  Equation (\ref{eq:mds_radius}), based on a constant rate of
  momentum injection (see below), describes the radial expansion after
  $\tsfm$ quite well (see Figures~\ref{fig:tevol_t001}-\ref{fig:tevol_t1}(a)).

\item {\it Morphology}

  Because of the highly inhomogeneous structure of the ``background''
  warm/cold ISM into which they propagate, SBs have complex morphology
  (Figures~\ref{fig:slice_n1_t001}-\ref{fig:zoom}).  Fingers and
  islands of hot, warm, and cold gas phases interpenetrate, with
  irregular interfaces.  Nevertheless, the SBs in our simulations
  retain the traditional elements of a very hot, very low density interior
  contained within a shell consisting of shocked, cooled, and
  compressed ambient gas. Except at the earliest stages, the expansion
  velocity of the hot medium exceeds that of the surrounding shell. In
  models with $\dtsn = 0.01,0.1\Myr$, the interior remains
  overpressured relative to the ambient ISM, whereas in models with
  $\dtsn=1\Myr$, the pressure can drop below ambient values at late
  time.  Pressures in the hot interior can also either be higher or lower than
  in the warm shell.
  SB interiors include dense clouds that were shock-heated and accelerated but
  left behind by the more rapid advance of the outer front; these clouds may
  remain warm if $\dtsn$ is sufficiently small, or they may cool back
  down if $\dtsn$ is large.

\item {\it Energetics of gas phases}

  For all of our models, the mean temperature $\Thot$ of the hot bubble
  interior remains $> 10^6 \Kel$ throughout the simulation.
  Figures~\ref{fig:tevol_t001} -- \ref{fig:tevol_t1} show that $\Thot$
  remains close to $10^7\Kel$ for models with $\dtsn=0.01\Myr$,
  evolving continuously when $\navg$ is low.  Models with higher
  $\dtsn$ and $\navg$ show spikes in $\Thot$ after each event.  PDFs in
  the temperature-velocity plane
  (Figure~\ref{fig:Tvpdf}) at $\tH$ show differences
  for models in the ``continuous'' ($\dtsn < \tsfm$) vs. ``discrete''
  ($\dtsn > \tsfm$) limit.  For the former, shocked dense clouds that
  are originally CNM are maintained at $T \sim 10^4\Kel$ by continuous
  heating; they are also accelerated up to a few tens of $\kms$
    (Figure~\ref{fig:nvpdf}).  For
  the latter, dense CNM clouds are shocked and accelerated up to $\sim
  10\kms$, but they cool back to $\sim 100\Kel$.  For all models, the
  SB shell is mostly composed of gas that was originally WNM before
  being shocked and swept up; it remains at $T\sim 10^4\Kel$, with
  supersonic velocities of several 10's to $> 100\kms$.  Most of the
  mass of warm gas has velocity below $100\kms$, so it would not be
  able to escape from the gravitational potential of a massive galaxy.
  However, substantial mass loss in warm gas would be expected for
  dwarf galaxies (see Figure~\ref{fig:vcum}).  For all
  cases except model \model{10}{1}, most of interior volume of the SB is
  filled by gas at $T\sim 10^7 - 10^8\Kel$.  Mass-weighted mean values
  at $\tH-\tHH$ are $\Thot=10^6-10^7\Kel$.
  Although the hot medium velocities exceed $\sim 100\kms$
  for all but models \model{10}{1} and \model{1}{1} (where $v_{\rm
    hot}$ is several 10s of $\kms$), the hot gas generally has
  enthalpy exceeding its kinetic energy and is at most transonic.
  Winds initiated with hot gas from SBs would accelerate as streamlines
  diverge after breakout, and have asymptotic velocities  up to $200-600\kms$.

\item {\it Momentum}

  Figure~\ref{fig:perSN_p} shows that for all
  models, $\phat$ remains relatively constant after $\tsfm$, in
  the range $ 0.7- 3\times 10^5\Msun \kms$.  That is, the SB evolves
  with nearly constant increase of momentum for each SN (or linear
  increase of momentum in time), quite different from the classical
  pressure-driven snowplow solution with constant increase of energy
  for each SN (linear increase of energy in time).
  Figures~\ref{fig:tevol_t001}-\ref{fig:tevol_t1}(e) show good agreement
  with $p_b=\phat t/\dtsn$.    The value of
  $\phat$ is very insensitive to the ambient density, and 
  increases slightly at higher $\dtsn$.  The values we obtain for
  $\phat$ are similar to the final momentum obtained in
  recent simulations of SNR expansion following a single SN explosion
  in an inhomogeneous medium
  \citep[KO15,]{2015A&A...576A..95I,2015MNRAS.450..504M,2015MNRAS.451.2757W},
  as well as for the homogeneous medium case with a single SN 
  \citep[][KO15]{1988ApJ...334..252C,1998ApJ...500..342B,1998ApJ...500...95T}.

  Recently, \citet{2016arXiv160601242G} have argued, based on
  spherically symmetric simulations of multiple SNe in
  a uniform background medium conducted with a Lagrangian code,
  that the mean momentum injection per SN to the ISM, $p_*$, may be
  higher for a SB than for an individual SNR.  Indeed, Equation
  (\ref{eq:sb_mom_1}) for the evolution prior to shell formation, or
  the same expression multiplied by 0.56 for the classical adiabatic
  pressure-driven snowplow, shows that if energy losses are small, the
  momentum per SN can exceed $10^6\Msun \kms$ at late times.  However,
  there are two difficulties in applying the results of
  \citet{2016arXiv160601242G} to the real ISM.  First, high values of
  the momentum/SN are achieved only at quite late times, beyond the
  point that the SB radius would have exceeded $H$.  Second, the
  extremely inhomogeneous conditions of the real ISM mean that a simple
  contact discontinuity between the hot interior and cooled shell
  cannot be maintained.  Instabilities initiated at interfaces (both
  with the shell and with embedded dense clouds) develop into
  turbulence, and the subsequent mixing between the hot medium and denser phases
  enhances cooling.  Spherically symmetric models cannot capture the
  energy losses that are inherent to evolution in a cloudy ISM.  While
  simulations at higher resolution than the present ones would be
  valuable to investigate the mixing and cooling at interfaces
  in greater detail,
  we find (see Appendix) that our results are converged.  This
  suggests that the high values of $p_*$ proposed
  by \citet{2016arXiv160601242G} would not apply in the real ISM.
  Indeed, within the context of models in which star formation rates
  are predominantly regulated by the momentum injection from SNe
  \citep[][]{2010ApJ...721..975O,2011ApJ...731...41O,2011ApJ...743...25K},
  a much larger value of $p_*$ would be inconsistent with observations
  of $\Sigma_{\rm SFR}$ in both normal galaxies and starbursts.

\item {\it Hot gas mass and wind loading}

  Figure~\ref{fig:perSN_hot} shows that  the hot gas mass per SN 
  peaks at a value $\Mhat\sim 400 - 2000 \Msun$ at $t \sim \tsfm$ and
  then drops.  For  most models, $\Mhat\sim 10 -100 \Msun$ for
  $t \sim \tH -\tHH$. The value of $\Mhat$ decreases for increasing
  background ISM density.  The late-time value of $\Mhat$ does
  not depend strongly on $\dtsn$, but because $\dtsn$ determines the
  time $\tH$ when a SB would begin to break out of the disk, the
  SN interval would affect the mass loading of winds by SBs.  Taking the
  wind hot gas mass loading $\betah = \Mhat(t)/100\Msun$ for $t\sim \tH - \tHH$,
  only our model \model{0.01}{0.01} has $\betah > 1$, and this is only
  for the first part of the ``breakout'' period.  We conclude that the
  potential for SBs to drive heavily mass-loaded hot winds depends strongly on
  $\dtsn$, or equivalently the mass of the star cluster driving the bubble.

  The time $\tH$ depends on the background ISM density and scale
  height, and Equation (\ref{eq:dt_hbo}) provides an expression for
  the maximum SN interval ($\dtsn < \Delta t_{\rm hbo}$) that would allow
  ``hot breakout,'' with the SB radius reaching $H$ prior to the onset of
  strong cooling ($\tH < \tsfm$).  The value $\Delta t_{\rm hbo}$ can be
  converted to a minimum cluster mass (or local mass of young stars)
  that enables hot breakout; Equation (\ref{eq:M_hbo}) gives this mass
  $M_{\rm cl, hbo}$
  as a function of local properties in the disk. Under typical
  galactic disk conditions, the condition for hot breakout would not
  be met.  This implies that $\betah <1$ would be expected for a hot wind driven
  by SBs for most regions in a galaxy.  However, starbursts in the centers
  of galaxies have very
  high local concentrations of young stars, often exceeding $M_{\rm cl,hbo}$.
  These are indeed exactly the systems where strong wind
  signatures are observed \citep[e.g.][]{2015ApJ...809..147H}.
  
  For dwarf galaxies with shallow potential wells, gas velocities need not
  reach hundreds of $\kms$ to escape as an outflow.  Except for our models
  with the $\dtsn = 1\Myr$ (which exceeds the expected mean local SN
  interval $\Delta t_{\rm SN,H} \sim 0.3 \Myr$), at $\tH$ there is more than
  $100 \Msun$ in mostly-warm gas per SN that has $|v_z| >  50 \kms$ 
  (see Figure~\ref{fig:vcum}).  This suggests that SBs could effectively
  clear the baryons from low mass halos, as is required to reconcile
  observed statistics of dwarfs with $\Lambda$CDM cosmology
  \citep[e.g.][]{2015ARA&A..53...51S}.
\end{enumerate}

Finally, we note that there are a number of physical effects that we
have not included in the present simulations, which potentially could
lead to substantial quantitative difference in some results.  In
particular, we have not incorporated thermal conduction, magnetic
fields, turbulence in background state, or a pre-existing hot phase,
all of which could alter the overall evolution and detailed density
and thermal structure of SBs.  Additionally, higher resolution would
aid in investigating the details of turbulent mixing at the interfaces
between phases.  Many of the above additional physical effects are
best addressed in fully self-consistent simulations of three-phase ISM
galactic disks with star formation and SNe, which we are currently
pursuing (C.-G. Kim \& E.C. Ostriker, in preparation).  Self-consistent
star-forming ISM disk simulations are also helpful in directly
measuring mass-loss rates in winds, without having to make an
assumption that SB properties when $r_b\sim H$ determine mass-loss rates
(in fact, our galactic disk ISM simulations show $\betah \sim 0.1-1$ in hot
gas, confirming the present results).  However, the isolation of individual
elements is extremely helpful in building deeper understanding of the
ISM, and we believe it will continue be fruitful to conduct focused simulations
and analyses of SBs, with enhanced physics and numerical resolution.

\acknowledgements

This work was supported by grant no. AST-1312006 from the National 
Science Foundation.  Simulations were performed on the computational
resources supported by the PICSciE TIGRESS High Performance Computing
Center at Princeton University.

\appendix

\section{Numerical Convergence}\label{sec:convergence}

In KO15, we showed that the evolution of a radiative SNR is 
numerically converged provided that 
the initial size of the feedback region is sufficiently small compared to
the shell formation radius, $r_{\rm init}/r_{\rm sf}<1/3$, and the resolution is high
enough to resolve the shell formation, $\Delta x/r_{\rm sf}<1/3$. Physically,
these criteria can be understood considering that all of the hot gas, and
most of the radial momentum, is produced  via 
propagation of very strong shocks during energy conserving stages
of evolution. In the post shell formation stage for an individual SNR,
some additional momentum
is acquired as the overpressured hot gas in the interior of the SNR
pushes the surrounding shell outward, but this effect is less significant 
than originally thought 
\citep[e.g.,][]{1977ApJ...218..148M,1988RvMP...60....1O}. 
Therefore, both momentum acquisition and hot gas creation can be numerically
converged if one resolves the energy conserving phase.

The evolution of a SB is different from that of a single SNR. It is still
important to resolve the onset of cooling in the shocked ambient medium,
with a physical scale described by the shell formation radius.
In principle, if $\dtsn$ is
sufficiently small, energy from subsequent SNe extends the energy-conserving
stage to $\tsfm > \tsf$ and produces a larger shell formation radius 
(see Equations \ref{eq:tsf} and \ref{eq:tsfm}).  This can in principle 
relax the resolution requirement for convergence, although 
in practice we still use the ``single SN'' criterion to set the
feedback region size for each individual
feedback event (see Section \ref{sec:method}).  

While early evolution of a single SNR and SB are similar, 
evolution after shell formation, and in particular the build-up of  
momentum and hot gas, is different for a SB from
either the energy-conserving or pressure-driven snowplow phase of a 
single SNR.  First, consider the case of a uniform ambient medium,
and neglect development of instabilities in the shell that
would lead to non-spherical morphology.  
After shell formation in a spherical SB,
if the SB has sufficiently low internal density,
ejecta from subsequent SNe would freely expand until reaching the dense shell.
In this case, as the ejecta hit the dense shell, a shock would run into 
the dense medium, and quickly cool down. At the same time, 
a reverse shock would propagate backward and heat up the interior.
If the density in the
interior of the SB is high enough for the ejecta to be slowed down
before reaching the shell, then a Sedov-like solution could develop
from forward and backward shock propagation, 
maintaining a hot and overpressured condition in the SB interior.   
If the SN interval is short enough, and thermalization of energy occurs
in such a way that the interior and shell are separated by a contact
discontinuity (i.e. without propagation of a shock into the shell, which
would then radiatively cool), evolution would follow the limit of 
classical SB evolution driven by continuous energy injection
\citep[e.g.,][]{1977ApJ...218..377W}.  Recent simulations 
 have followed SB evolution with cooling for a
uniform ambient medium under the assumption
that energy is fully thermalized at small scales;
\citet{2016arXiv160601242G}
impose spherical symmetry and use a Lagrangian code
to aid in resolving the interface between the SB interior and dense shell,
while \citet{2016arXiv160300815Y}
conduct fully three-dimensional simulations resolving down to $\sim 1 \pc$,
showing evolution that agrees with corresponding spherical models.

Unlike the idealized 1D spherical theory (or simulations) for a uniform
ambient medium, even in the limit of short $\dtsn$ that approaches continuous
energy injection, the evolution of a SB in the real ISM will be more complex.
Multi-dimensionality allows instabilities to develop at the interface with
the shocked cooled outer shell and internal overdense clumps that are an
inherent aspect of the warm/cold ISM.  These instabilities result in
hydrodynamic mixing between phases, and enhance cooling.
If thermal conduction is considered, the mass and energy exchanges
between hot interior and cooled shell will also be enhanced.
Especially considering the role of turbulence (driven by instabilities) in
creating structure and mixing material at fine scales, the numerical
requirements needed to capture the impact of multiple SN explosions in
a cloudy ISM are not obvious -- and indeed the numerical requirements
may differ, depending on what issue is in question.  
Numerical simulations with grid resolution of order of parsec 
cannot resolve the realistic Field length \citep{1990ApJ...358..375B},
so that the total cooling is dominated by unresolved interfaces.
In spherical symmetry, one might expect the total cooling rate to vary
$\propto \rbub^2 \Delta x $, so that for a given shell size cooling would be
overestimated at lower resolution.  Also, with a clumpy medium, the usual
realization of SN feedback with purely thermal energy is in question.

\begin{figure}
\plotone{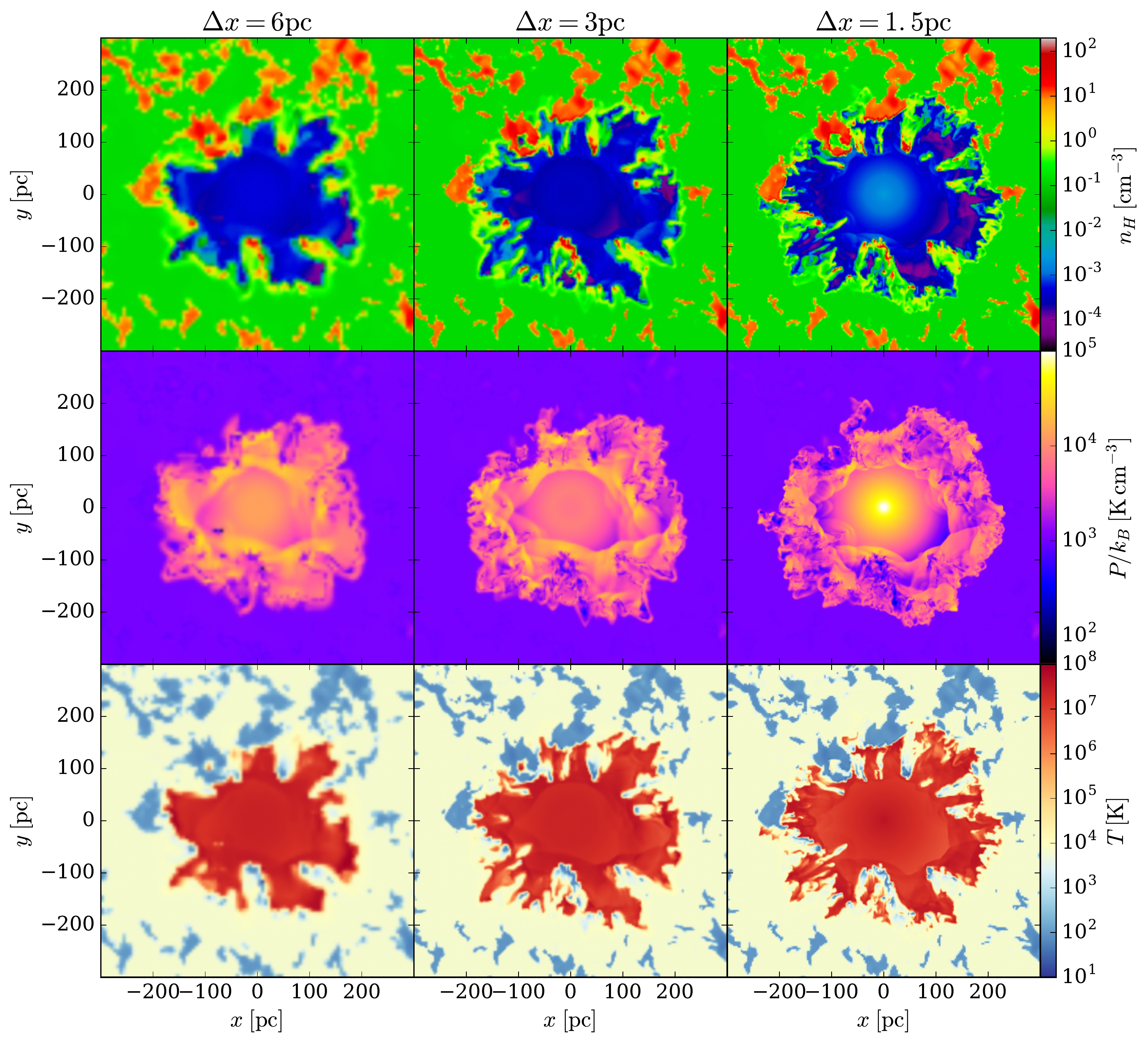}
\caption{
Slices at $t=4\Myr$ for low (left), standard (middle), and high (right)
resolution simulations of Model \model{1}{0.1}.
\label{fig:convergence}}
\end{figure}

\begin{figure}
\plotone{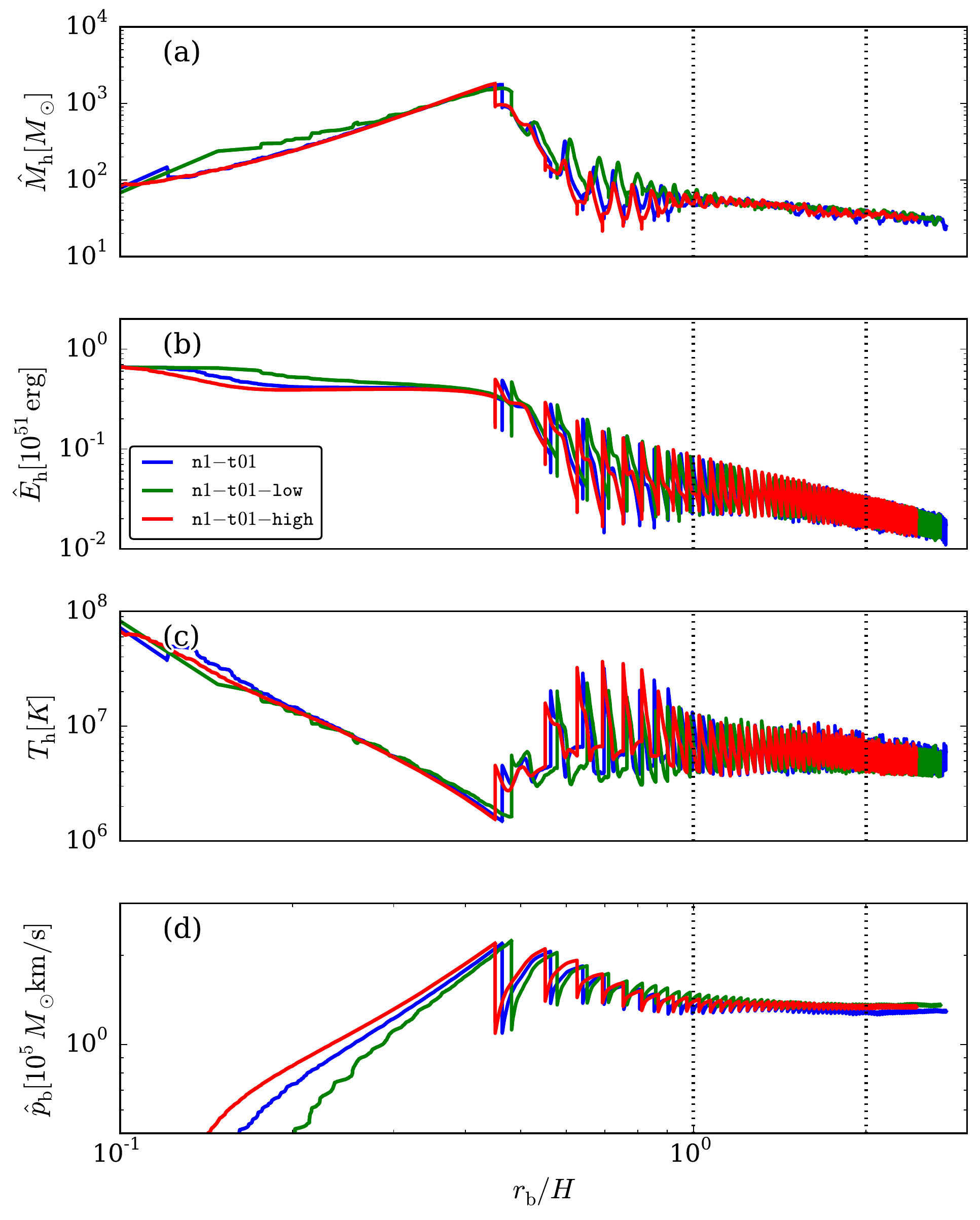}
\caption{Resolution study of Model \model{1}{0.1}.  Panels show per-SN values
of  (a) hot gas mass, (b) hot gas thermal energy, and (d) bubble momentum, as
well as (c) the mean temperature of the hot component.  Blue, green, and red
lines denote resolution $\Delta x=6\pc$, $3\pc$, and $1.5\pc$, respectively.
\label{fig:perSN_res}}
\end{figure}

In order to address these concerns, we perform two numerical convergence 
tests. First, we conduct a resolution test by re-running 
Model \model{1}{0.1} with a factor of two higher and lower resolutions, 
\model{1}{0.1-high} and \model{1}{0.1-low}, respectively.
In order to keep the background state for different resolutions, we adopt
the same initial condition from the saturated state of thermal instability
simulations with standard $3\pc$ resolution and then 
refine/degrade for different resolutions.
Figure~\ref{fig:convergence} illustrates the difference in structure at
$t=4\Myr$ for different resolutions.    
In Figure~\ref{fig:perSN_res}, we plot all key quantities as a function of
normalized size of bubble $\rbub/H$: (a) hot gas mass per SN $\Mhat$, 
(b) hot gas thermal energy per SN $\Ehat$, 
(c) mass-weighted mean temperature of the hot gas $\Thot$,
and (d) bubble radial momentum per SN event $\phat$. 
The detailed evolution is slightly shifted toward the left for higher
resolution simulation. This means that the evolution is slightly faster
at higher resolution. However, the results for mass, energy, and momentum
loading, and for the mean interior temperature of the SB, are in agreement at
all resolutions, indicating that these integrated quantities are converged.

\begin{figure}
\plotone{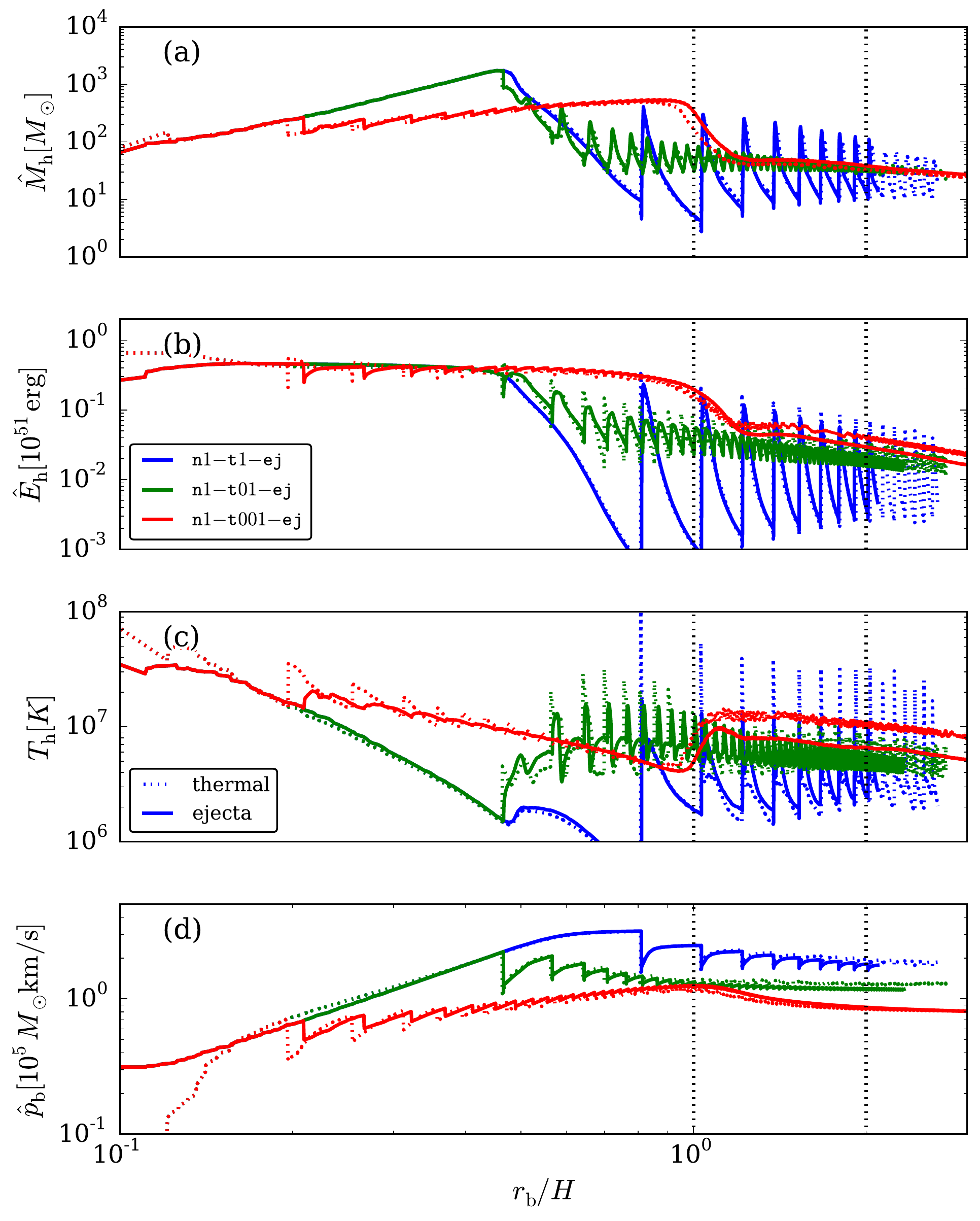}
\caption{Convergence study for two feedback realizations, ``ejecta'' (solid)
and ``thermal'' (dotted) feedback. Panels show per-SN values
of  (a) hot gas mass, (b) hot gas thermal energy, and (d) bubble momentum, as
well as (c) the mean temperature of the hot component.  Blue, green, and red
lines denote Models \model{1}{1}, \model{1}{0.1}, and \model{1}{0.01},
 respectively.
\label{fig:perSN_ej}}
\end{figure}

Second, we conduct a test with a different realization of SN feedback.  Instead
of using pure thermal energy (``thermal'' feedback), 
we dump ejecta mass $10\Msun$ and pure kinetic energy within a region
that encloses ambient medium mass not exceeding 
10\% of the ejecta mass (``ejecta'' feedback).
Figure~\ref{fig:perSN_ej} plots the same key quantities as in 
Figure~\ref{fig:perSN_res}. We plot results using
``ejecta'' feedback as solid lines
and results using the standard ``thermal'' feedback 
as dotted lines for Models \model{1}{1} (blue),
 \model{1}{0.1} (green), and \model{1}{0.01} (red). Again, there are
 small detailed differences, but the final results are generally in agreement
 for the two feedback treatments.  
In (b) and (c), the hot gas energy and temperature are slightly
lower in \model{1}{0.01-ej} than in \model{1}{0.01} since thermalization of the ejecta is 
not perfect when $\dtsn$ is short. However, the hot gas mass (in (a))
is consistent for the two feedback treatments,
implying that the main contributor to new hot gas is not the ejecta
but shock-heated existing gas in the SB interior.  
From examining the detailed evolution of both models, we clearly observe
develompent of a shock that propagates through the hot interior and
hits the CNM and WNM in the shell and fingers, generating new hot gas.
As a consequence, the ejecta
mass we use here also do not affect the results (unless it is too large).
The injected momentum is slightly decreased (less than 10\%) in higher
density models with ejecta feedback compared to thermal feedback.

\bibliographystyle{aasjournal} 
\bibliography{ms.bbl}

\end{document}